\documentclass[%
 aip,
 pof,
 amsmath,amssymb,
 reprint,%
]{revtex4-2} % revtex4-2 is used instead of revtex4-1, as the latter is outdated and does not exist anymore

\usepackage{blindtext}
\usepackage{graphicx}
\usepackage{dcolumn}% Align table columns on decimal point
\usepackage{bm}% bold math
\usepackage[utf8]{inputenc}
\usepackage[T1]{fontenc}
\usepackage{mathptmx}
\usepackage{etoolbox}
\usepackage{upgreek} % for upright greek letters
\usepackage[normalem]{ulem} % for strikeout
\usepackage{cancel} % for strikeout of equations

\usepackage{amsmath}
\usepackage{amsthm}
\usepackage{amssymb}
\usepackage{esint}
\usepackage[linkcolor=blue,colorlinks=true]{hyperref}
\usepackage{multirow}
\usepackage{mathrsfs} % for mathscr font

\makeatletter % better use internal names

\newenvironment{newnumbering*}[1][Alph]
 {%
  \def\jr@counter{#1}%
  \jr@setup@numbering{#1}{0}%
 }
 {%
  \setcounter{jr@\jr@counter @equation}{\value{equation}}%
  \setcounter{equation}{\value{jr@equation@save}}%
  \ignorespacesafterend
 }

\newcounter{jr@equation@save}
\newcommand{\jr@setup@numbering}[2]{%
  \@ifundefined{c@jr@#1@equation}{\newcounter{jr@#1@equation}}{}%
  \setcounter{jr@equation@save}{\value{equation}}%
  \setcounter{equation}{%
    \ifnum#2>0
      \value{jr@#1@equation}%
    \else
      0%
    \fi
  }%
  \renewcommand{\theequation}{\csname#1\endcsname{equation}}%
}
\makeatother

\makeatletter
\def\@email#1#2{%
 \endgroup
 \patchcmd{\titleblock@produce}
  {\frontmatter@RRAPformat}
  {\frontmatter@RRAPformat{\produce@RRAP{*#1\href{mailto:#2}{#2}}}\frontmatter@RRAPformat}
  {}{}
}%
\makeatother

\begin{document}

\title[Simulation of the High Mach Number Motion for Bubble Collapse in a Compressible Euler Fluid using Basilisk]{Simulation of the High Mach Number Motion for Bubble Collapse in a Compressible Euler Fluid using Basilisk}
\author{Daniels Krimans}
\email{krimans@physik.uni-kiel.de}
\affiliation{Institute of Theoretical Physics and Astrophysics, Christian-Albrechts-Universität zu Kiel, 24098 Kiel, Germany}
\author{Steven J. Ruuth}
\affiliation{Department of Mathematics, Simon Fraser University, Burnaby, British Columbia V5A 1S6, Canada}
\author{Seth Putterman}
\affiliation{Physics and Astronomy Department, University of California Los Angeles, Los Angeles, California 90095, USA}

\date{21 May 2025}

\begin{abstract} 
We examine an extreme case of experimentally realizable sonoluminescence, where spherical cavities have an initial radius that is $10$ to $20$ times their ambient radius and change their radius by a factor of over $100$ during the collapse.
Among the many physical processes at play, we focus on fluid compressibility, modeled using the Tait-Murnaghan equation of state for a homentropic Euler fluid.
To capture such extreme motion, with Mach numbers relative to ambient sound speed greater than one during the final stages of implosion, requires methods beyond the classic approaches of Rayleigh and Gilmore.
In this direction, we applied an all-Mach solver developed in the Basilisk framework, actively used to model bubble dynamics.
To capture high Mach number motion and resolve dynamics in the sonoluminescence regime, we employed the well-established uniform bubble approximation for the ideal gas inside the bubble.
Within this approximation, the all-Mach solver achieved numerically converging results describing the evolution of the bubble wall $R(t)$.
Although compressibility slows down the collapse, these bubbles reach velocities exceeding the ambient speed of sound of the surrounding fluid.
Our method works for various fluids and is applied to liquid lithium as well as water.
Our results reproduce the equation-of-state-dependent asymptotic power-law region predicted by analytic calculations for water and liquid lithium in the case of an empty cavity.
When the cavity is filled with an ideal gas, the transition to Mach number greater than one in liquid lithium occurs later in the collapse than for water, making liquid lithium a possible candidate for achieving greater concentration of energy density.
Furthermore, an outgoing shock wave, which can diagnose cavitation in opaque fluids such as liquid lithium, is captured without implementing an ad hoc construction algorithm.
\end{abstract}

\maketitle

%%%%%%%%%%%%%%%%%%%%%%%%%%%%%%%%%%%%%%%%%%%%%%%%%%%%%%%%%%%%%%%%%%%%%%%%%%%%%%%%%%%%%%%%%%%%%%%%%%%%%%%%%%%%%%%%%%%%%%%%%
% 
%                      INTRODUCTION
%
%%%%%%%%%%%%%%%%%%%%%%%%%%%%%%%%%%%%%%%%%%%%%%%%%%%%%%%%%%%%%%%%%%%%%%%%%%%%%%%%%%%%%%%%%%%%%%%%%%%%%%%%%%%%%%%%%%%%%%%%%

\section{Introduction}

Cavitation is the phenomenon where bubbles spontaneously form in a moving fluid.
Their subsequent motion can be so strong as to damage nearby boundaries \cite{annurev:/content/journals/10.1146/annurev.fl.19.010187.000531, KUMAR2010374}, and their implosion can even form a light-emitting plasma, a process known as sonoluminescence \cite{doi:10.1080/00018738400101711, BARBER199765}.
The singularity underlying this multiscale process was derived by Rayleigh \cite{doi:10.1080/14786440808635681} for an empty cavity surrounded by an incompressible fluid.
He showed that, for spherical symmetry, at the late stages of the collapse, the radius collapses to zero at a finite time $t_c$ according to
\begin{equation}
\label{incompressible_Rayleigh_R_t}
\begin{gathered}
R(t) = A(t_c - t)^n,
\end{gathered}
\end{equation}
where $n = 2/5$, $A = (5/2)^{2/5} (E/2 \pi \rho_{\infty})^{1/5}$, $\rho_{\infty}$ is the ambient mass density of the surrounding fluid at an external pressure $p_{\infty}$, and $E = (4\pi/3)p_{\infty} R_m^3$ is the initial energy of the fluid, which is the work required to form a cavity of radius $R_m$.
If the cavity contains gas, this implosion can lead to a strong concentration of energy density, turning the gas into plasma, which emits a picosecond-long flash of light \cite{doi.org/10.1038/352318a0}.
Even when the cavity is empty, so that the pressure at the cavity wall is zero, at the late stages of the collapse a large pressure gradient appears inside the fluid near the wall, given by \cite{doi:10.1080/14786440808635681, Batchelor_2000}
\begin{equation}
\label{incompressible_pressure_empty_cavity}
\begin{gathered}
p(r,t) = \frac{\rho_{\infty}\dot{R}^2(t)}{2} \left( \frac{R(t)}{r} - \left( \frac{R(t)}{r} \right)^4 \right),
\end{gathered}
\end{equation}
where $\dot{R}(t)$ is the velocity of the cavity wall, which can be computed using Eq.~\eqref{incompressible_Rayleigh_R_t}.
\par

The maximum pressure in the fluid is $p_{\textrm{max}} = 0.16 p_{\infty} (R_m/R)^3$ and occurs at $r = 1.6R$, where $R$ is the instantaneous radius of the cavity.
For $R_m/R = 27$ and $p_{\infty} = 1\textrm{ atm}$, the pressure within the fluid exceeds $3000\textrm{ atm}$, which is the yield stress of water. 
Here, we refer to the Tait-Murnaghan equation of state
\begin{equation}
\label{Tait_Murnaghan_EoS}
\begin{gathered}
p(\rho) = B \left( \left( \frac{\rho}{\rho_0} \right)^{\Gamma} - 1 \right),
\end{gathered}
\end{equation}
which will be assumed for all compressible fluids under consideration, where $\Gamma$ is a constant, but both $B$ and $\rho_0$ are slowly varying functions of specific entropy \cite{Cole_1948, Hunter_1960, Batchelor_2000}.
For our purposes, we treat $B$ and $\rho_0$ as constants as well.
In this equation, $\rho_0$ is the density at zero pressure, and for water \cite{Cole_1948, Hunter_1960, Batchelor_2000}, $\Gamma = 7$.
Since typically $B \gg 1\textrm{ atm}$, $\rho_0$ can be estimated using the known value of the density at $p = 1\textrm{ atm}$.
For example, taking the density of water at $p = 1\textrm{ atm}$ and temperature $T = 20\textrm{ }^{\circ}\textrm{C}$ from Ref.~\onlinecite{10.1063/1.1461829} yields $\rho_0 = 998\textrm{ kg/}\mathrm{m^3}$.
\par

The adiabatic speed of sound, $c$, for such an equation of state is given by
\begin{equation}
\label{Tait_Murnaghan_speed_of_sound}
\begin{gathered}
c^2 = \frac{\Gamma (p + B)}{\rho},
\end{gathered}
\end{equation}
which also allows us to define the ambient speed of sound, $c_{\infty}$, by evaluating at pressure $p_{\infty}$ and density $\rho_{\infty}$.
This is used to define the Mach number for a collapse in a compressible fluid as $\textrm{M} = \dot{R}/c_{\infty}$.
Additionally, experimental data on the speed of sound is used to estimate values of $B$ for various fluids.
For example, for water \cite{10.1063/1.1461829}, $c = 1.48 \times 10^3\textrm{ m/s}$ at $p = 1\textrm{ atm}$ and temperature $T = 20\textrm{ }^{\circ}\textrm{C}$, yielding $B = 3.12 \times 10^8\textrm{ Pa}$.
\par

If cavitation occurs in water, compressibility needs to be considered when the implosions are strong enough that $R_m / R > 27$ is reached. 
In addition to its maximum radius, a sonoluminescing bubble is characterized by its ambient radius, $R_0$, when the gas pressure equals $p_{\infty}$, and its minimum radius, $a$, which is largely determined by the van der Waals hard core size of the gas atoms. 
This phenomenon spans a large parameter space: typically $7 < R_0 / a < 10$, while $8 < R_m / R_0 < 24$ has been achieved for light-emitting bubbles \cite{PhysRevLett.85.3037, PhysRevLett.101.234301}.
Taken together, typical sonoluminescent bubbles in water reach $R_m / R > 100$, so, as emphasized by Ramsey \cite{PhysRevLett.110.154301}, compressibility must be considered if one is interested in probing the extent to which cavitation focuses energy.
\par

The speed at which a cavity implodes in an incompressible fluid after the initial stages of collapse is given by the following, as derived from Eq.~\eqref{incompressible_Rayleigh_R_t}:
\begin{equation}
\label{incompressible_Rayleigh_velocity}
\begin{gathered}
\dot{R}(t) = -\sqrt{\frac{2p_{\infty}}{3\rho_{\infty}}} \left( \frac{R_m}{R(t)} \right)^{3/2}
\end{gathered}
\end{equation}
Comparing this velocity to the ambient speed of sound in water, $c_{\infty} = 1.48 \times 10^3 \textrm{ m/s}$, obtained using the values $p_{\infty} = 1\textrm{ atm}$ and $\rho_{\infty} = 998\textrm{ kg/}\mathrm{m^3}$, yields a Mach number $\textrm{M} = 0.78$ when $R_m/R = 27$.
Therefore, describing cavitation collapse in a real fluid requires the analysis of high Mach number motion.
This point was made by Hunter \cite{Hunter_1960}, whose asymptotic calculation for the final stage of collapse of an empty cavity, including the effects of compressibility of the surrounding fluid, yielded a power-law solution of the form given in Eq.~\eqref{incompressible_Rayleigh_R_t} with $n=0.56$ for water.
His approach has been extended to a wide range of fluids with various values of $\Gamma$ \cite{10.1063/5.0160469}.
\par

We will also compare the theory to fluids with different compressibility than water, but in which cavitation has also been studied. 
One such example is liquid lithium \cite{10.1115/1.3609699, PhysRevC.85.054620}, for which the parameters in the equation of state at its melting temperature $T = 454\textrm{ K}$ and $p = 1\textrm{ atm}$ are estimated from Ref.~\onlinecite{Davison1968CompilationOT} to be $\rho_0 = 516\textrm{ kg/}\mathrm{m^3}$ and from Ref.~\onlinecite{doi:10.1179/174328007X212490} to be $c = 4.54 \times 10^3\textrm{ m/s}$.
We estimate $\Gamma$ and $B$ based on Fig.~3 of Ref.~\onlinecite{Khishchenko_2018}, where $\rho(p = 10\textrm{ GPa})/\rho_0 = 1.5$.
Combining this with the speed of sound, Eqs.~\eqref{Tait_Murnaghan_EoS} and \eqref{Tait_Murnaghan_speed_of_sound} can be solved simultaneously, yielding $\Gamma = 3.75$ and $B = 2.83 \times 10^9\textrm{ Pa}$.
We retain three significant figures for the calculations, though the experimental data for these parameters are not that accurate.
The data from Ref.~\onlinecite{doi:10.1021/acs.jpcb.0c05324} yields a consistent value of $\Gamma$ when the cube of the wavenumber of the first peak of the structure factor, $q$, is interpreted as yielding the density at the given pressure by $\rho(p)/\rho_0 = (q(p) / q_0 )^3$, where the subscript ``0'' refers to zero pressure.
In liquid lithium \cite{10.1063/5.0160469}, the power-law solution at the final stage of collapse gives an exponent of $n = 0.65$.
\par

The initial stages of bubble collapse can be well described by the incompressible Euler equations.
As $R$ decreases, there is a transition to the asymptotic regime determined by $n(\Gamma)$.
Understanding the evolution of energy density requires linking the initial state to the final stages of implosion, which in turn necessitates a description of the transition region and a solution to the Euler equations for all Mach numbers.
The goal of this paper is to demonstrate such a transition for the case of spherical symmetry. 
To achieve this, we follow Hunter \cite{Hunter_1960} and assume homentropic flow, meaning the specific entropy is uniform and constant over time.
This simplifies the calculations, as a simple equation of state given by Eq.~\eqref{Tait_Murnaghan_EoS} can then be used, for which analytic asymptotic results are known.
This allows us to confirm the accuracy of the numerical solution obtained for the final stages of the collapse.
Moreover, we believe that for the rapid collapses of interest, ignoring heat transfer is a reasonable approximation.
For further simplicity, we also neglect the effects of viscosity, surface tension, and mass transfer.
\par

Although compressible cavitation has a long history, it remains a topic of intense current research.
Gilmore \cite{gilmore_2023_s7hvs-xmw23} used the method of characteristics, implemented with the Kirkwood-Bethe ansatz, to approximately calculate compressible corrections to the incompressible Rayleigh-Plesset equation for small Mach numbers.
He found an asymptotic limit of $n = 2/3$, compared to $n = 2/5$ for Rayleigh and $n=0.56$ as found by Hunter.
Hickling and Plesset \cite{10.1063/1.1711058} solved the homentropic compressible Euler equations for the fluid outside an ideal gas cavity using both the Lagrangian variables and the characteristic form.
In the case of an empty cavity, they confirmed the value $n=0.56$ and found a different collapse radius from Gilmore, corresponding to the difference in $n$.
\par 

Shock wave emission from strong bubble collapse is another key aspect of sonoluminescence \cite{PhysRevLett.84.1328, PhysRevE.61.R1020}.
For experiments on opaque fluids, such as liquid lithium, shocks provide a key diagnostic.
Numerical methods used by Hickling and Plesset predict shock wave formation but do not capture the subsequent motion.
As Denner and Schenke note \cite{10.1063/5.0131930}, characteristics will eventually cross, leading to a multivalued solution, which requires supplementing the equations with an ad hoc hypothesis.
Koch et al.\cite{KOCH201671} also noted differences with Gilmore for higher Mach numbers.
Nevertheless, the remarkable accuracy of the Gilmore equations has led to their use in describing bubble collapse and shock wave emission \cite{Lauterborn_2010, Liang_Linz_Freidank_Paltauf_Vogel_2022, 10.1063/5.0131930}.
\par

To gain insight into sonoluminescence, Wu and Roberts \cite{PhysRevLett.70.3424} solved the Euler equations for the motion of gas within the bubble.
The motion of the cavity wall was determined by Rayleigh’s differential equation for incompressible flow, modified to include sound radiation from the cavity wall.
They found that an imploding shock wave could form inside the bubble.
Our ultimate goal is to generalize this picture to the case where the external fluid is compressible and the interior of the bubble is described via molecular dynamics.
However, as a first step, we aim to unify compressibility and shock emission in the external fluid without the need for additional inputs, focusing on the case where the gas inside the bubble is described by hydrodynamics, particularly for extreme bubble implosions.
Fuster, Dopazo, and Hauke \cite{10.1121/1.3502464} have already computed such spherical bubble collapse by solving full hydrodynamic equations for both the gas inside the cavity and the compressible fluid outside it, considering collapses where $R_m/R = 13$ was reached.
Successes have also been achieved using OpenFOAM software \cite{KOCH201671}, but it requires adjustments for mass conservation \cite{MILLER2013132}.
Impressive results have been achieved by Nguyen et al. \cite{10.1063/5.0167338} for extreme bubble motion, where the vapor plays a key role in the motion. 
Their results should reach the asymptotic regime \cite{Hunter_1960, 10.1063/5.0160469}, but this has not yet been conclusively determined.
\par

To approach this problem from a new perspective, we employed the readily available and actively used all-Mach method of Fuster and Popinet \cite{FUSTER2018752}, implemented in the Basilisk framework using the volume-of-fluid method.
Besides optimizing computational time, their approach also provides the capability for capturing shock waves, including surface tension, and aspherical response \cite{FUSTER2018752}, as well as extension to transport processes \cite{SAADE2023111865}.
In particular, this method was used to solve the problem without assuming spherical symmetry \cite{FUSTER2018752} and reached $R_m/R = 5$.
This formulation was later modified for spherical symmetry, and adiabatic collapses reaching $R_m/R = 13$ were analyzed \cite{SAADE2023111865}.
Nevertheless, these simulations did not reach sufficiently high values of $R_m/R$ and did not achieve the asymptotic limit \cite{Hunter_1960, 10.1063/5.0160469}.
\par

To simulate a homentropic, spherically collapsing cavity with a gas inside it, for $R_m/R > 100$, which is relevant for experiments, we first considered the mentioned all-Mach solvers \cite{FUSTER2018752,SAADE2023111865}.
Due to the extreme physical parameters encountered in such strong collapses, a numerical issue arises in which a region forms in the gas near the cavity wall with extremely low density and extremely high speed of sound. 
This occurs during the initial stages of the collapse, where incompressibility is a good approximation for the motion. 
The issue persisted even after refining the calculation to the limits of the available computational resources. 
\par

To resolve the issue, we modified the solver by introducing the uniform bubble approximation.
Instead of directly solving the Euler equations for the gas in the cavity, we assume that the pressure and the mass density of the gas are always uniform but time-dependent and consistent with the equation of state, which is taken to be an ideal adiabatic monatomic gas for simplicity. 
This approximation is accurate for early times, when the motion is nearly incompressible \cite{10.1063/1.858700}.
The uniform bubble approximation, though with different equations of state for the gas, is regularly used to describe physical processes driven by cavity collapse in incompressible fluids\cite{10.1063/1.858700, HILGENFELDT_BRENNER_GROSSMANN_LOHSE_1998}.
This approximation provides leading-order insight into the processes involved with energy focusing in cavitation.
It avoids the mentioned numerical artifacts and reaches regimes of motion where fluid compressibility is important.
However, it neglects later-time behavior that might be significant, such as shock waves forming in the gas.
Nevertheless, it offers a way to explore the effects of compressibility for large collapses and allows for quick and accurate computations.
\par

\begin{figure}
\includegraphics[width=4.8in, height=2.4in]{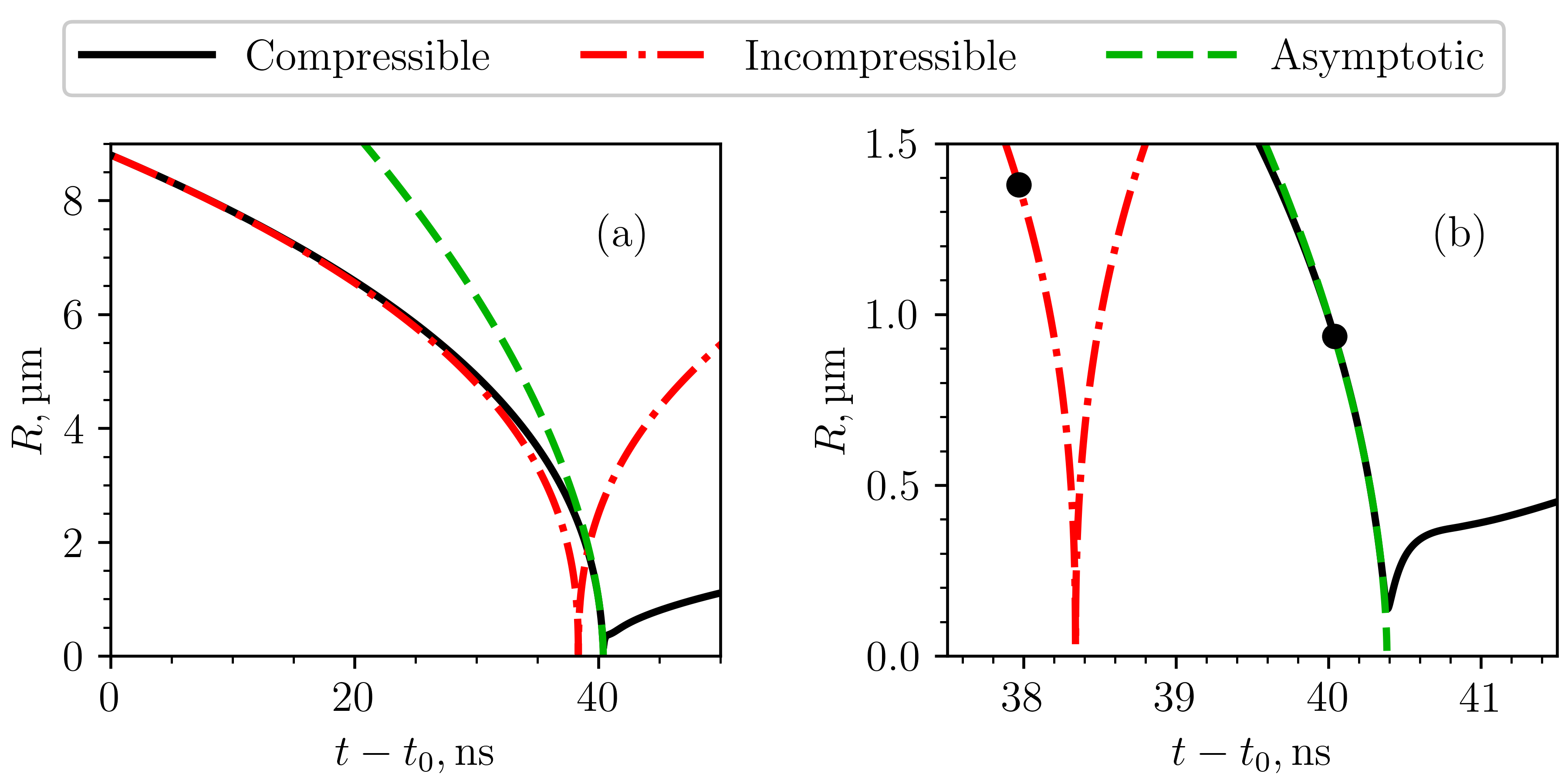} %for two-column figure, 6.69 inches x 8.25 inches. 
\caption{
Implosion of an ideal gas xenon bubble in water at ambient pressure $p_{\infty} = 1\textrm{ atm}$ and ambient temperature $T_{\infty} = 293\textrm{ K}$, with $R_m/R_0 = 20$ and $R_0 = 2.2\textrm{ }\upmu\textrm{m}$.
The red dot-dashed line represents the incompressible solution for the ambient density of water, $\rho_{\infty} = 998\textrm{ kg/}\mathrm{m^3}$.
The black solid line is obtained using our modified all-Mach solver with the uniform bubble approximation.
The compressibility of water is given by Eq.~\eqref{Tait_Murnaghan_EoS} with $\rho_0 = 998\textrm{ kg/}\mathrm{m^3}$, $\Gamma = 7$, and $B = 3.12 \times 10^8\textrm{ Pa}$.
The black dots indicate the point at which the Mach number $\textrm{M} = \dot{R}/c_{\infty} = 1$ relative to water is reached.
The dashed green line is a power-law fit of the form as in Eq.~\eqref{incompressible_Rayleigh_R_t} with $n = 0.56$, as predicted in Ref.~\onlinecite{Hunter_1960}.
The comparison is shown from time $t_0$, where $R(t_0) = R_m/5$, as before this time the compressible and incompressible results agree.
In (b), the solutions shown in (a) are zoomed in near the minimum radii of the curves. 
}
\label{fig:comparison_of_water}
\end{figure}

Fig.~\ref{fig:comparison_of_water} shows plots of the transition to the asymptotic domain of a compressible implosion in water.
For $R_m/R_0 = 20$ and $R_0 = 2.2\textrm{ }\upmu\textrm{m}$, motivated by experiments \cite{PhysRevLett.101.234301} for a cavity with ambient pressure $p_{\infty} = 1\textrm{ atm}$, ambient temperature $T_{\infty} = 293\textrm{ K}$, and here containing an ideal xenon gas, the result based on our implementation of the all-Mach solver is shown as the black solid line.
The maximum Mach number, $\textrm{M} = \dot{R}/c_{\infty}$, in this calculation is $2.7$, and the maximum ratio of radii is $R_m/R = 319$.
This highlights the necessity of an all-Mach solver to include compressibility effects for strong collapses observed in experiments.
The red dot-dashed lines in Fig.~\ref{fig:comparison_of_water} represent solutions describing the incompressible limit \cite{Batchelor_2000}.
If the water remained incompressible, the Mach number would reach $134$ and the maximum ratio of radii would be $R_m/R = 1460$.
Fig.~\ref{fig:comparison_of_water} illustrates the goal of this paper, which is to demonstrate the transition to the compressibility asymptote that controls the moment of collapse.
The green dashed curve corresponds to the power-law solution of the form given in Eq.~\eqref{incompressible_Rayleigh_R_t} with Hunter’s value $n = 0.56$, which is a good fit for the asymptotic motion that precedes the moment of collapse. 
The precise discretization parameters and details of the computations used to obtain the results are discussed in Section~\ref{subsection_20_to_1_collapses}.
\par

\par

\begin{figure}
\includegraphics[width=4.8in, height=2.4in]{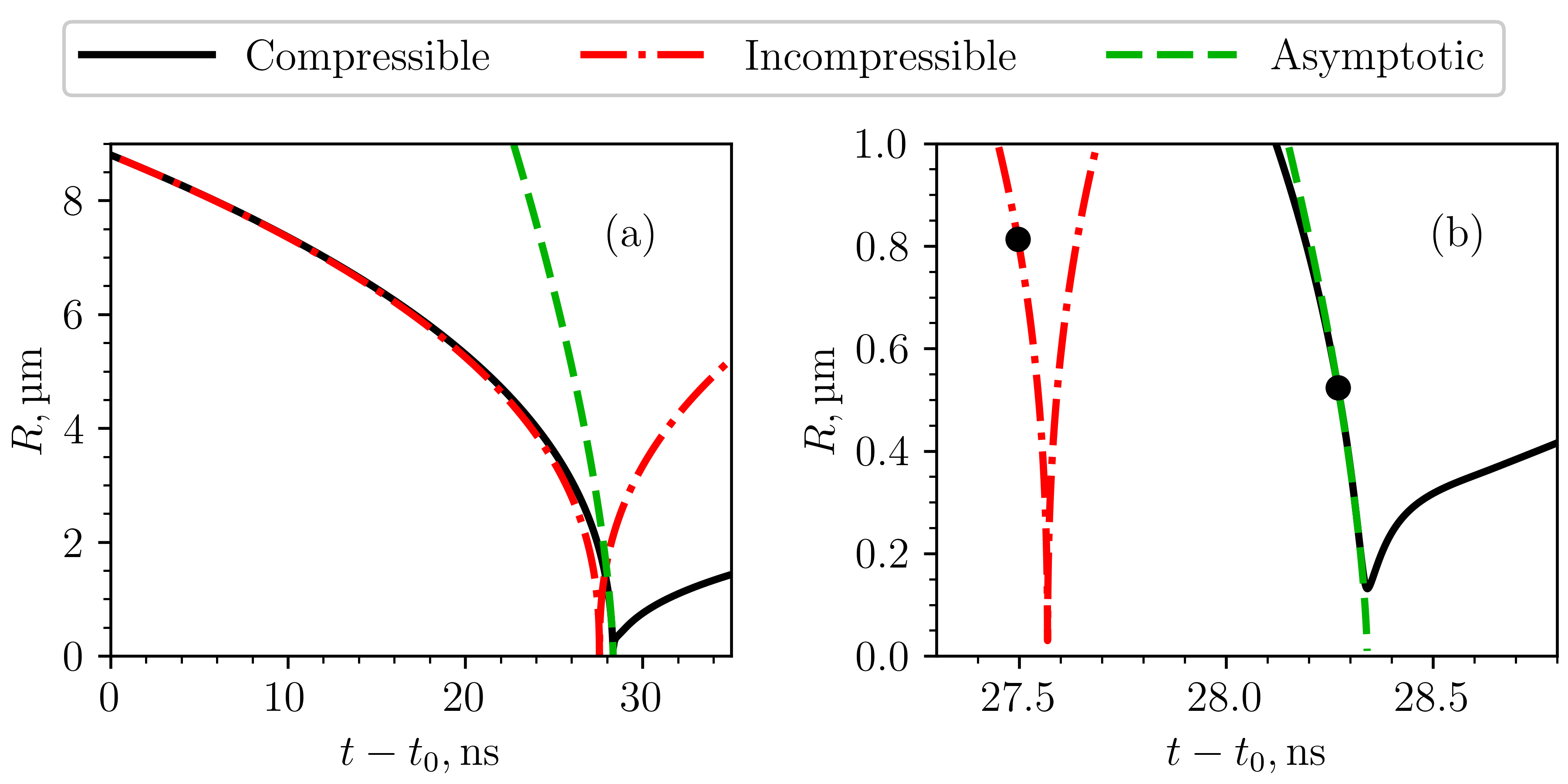} %for two-column figure, 6.69 inches x 8.25 inches. 
\caption{
Implosion of an ideal gas xenon bubble in liquid lithium at ambient pressure $p_{\infty} = 1\textrm{ atm}$ and ambient temperature $T_{\infty} = 454\textrm{ K}$, with $R_m/R_0 = 20$ and $R_0 = 2.2\textrm{ }\upmu\textrm{m}$.
The red dot-dashed line represents the incompressible solution for the ambient density of liquid lithium, $\rho_{\infty} = 516\textrm{ kg/}\mathrm{m^3}$.
The black solid line is obtained using our modified all-Mach solver with the uniform bubble approximation.
The compressibility of liquid lithium is given by Eq.~\eqref{Tait_Murnaghan_EoS} with $\rho_0 = 516\textrm{ kg/}\mathrm{m^3}$, $\Gamma = 3.75$, and $B = 2.83 \times 10^9\textrm{ Pa}$.
The black dots indicate the point at which the Mach number $\textrm{M} = \dot{R}/c_{\infty} = 1$ relative to liquid lithium is reached.
The dashed green line is a power-law fit of the form as in Eq.~\eqref{incompressible_Rayleigh_R_t} with $n = 0.65$, as predicted in Ref.~\onlinecite{10.1063/5.0160469}.
The comparison is shown from time $t_0$, where $R(t_0) = R_m/5$, as before this time the compressible and incompressible results agree.
In (b), the solutions shown in (a) are zoomed in near the minimum radii of the curves. 
}
\label{fig:comparison_of_lithium}
\end{figure}

Cavitation in liquid lithium has been studied in relation to erosion \cite{10.1115/1.3609699} and nuclear processes \cite{PhysRevC.85.054620}.
Using an ambient temperature of $T_{\infty} = 454\textrm{ K}$ relevant for the experiments, the motion of a bubble with assumed parameters $R_m/R_0 = 20$ and $R_0 = 2.2\textrm{ }\upmu\textrm{m}$ is shown in Fig.~\ref{fig:comparison_of_lithium}.
We again assume that the cavity contains ideal xenon gas and the ambient pressure is $p_{\infty} = 1\textrm{ atm}$.
At small radii, the motion agrees with the power-law solution of the form given in Eq.~\eqref{incompressible_Rayleigh_R_t} with $n = 0.65$, consistent with the asymptotic result \cite{10.1063/5.0160469}.
It can be noted that in this case, compared to water, the asymptotic regime is achieved at much smaller radii.
This can likely be explained by the larger value of $B$ for liquid lithium, which causes the motion to remain incompressible for a longer time.
The maximum Mach number reached with respect to liquid lithium, $\textrm{M} = \dot{R}/c_{\infty}$, is $61$ for the incompressible case, whereas our compressible calculation reaches $1.6$.
This corresponds to a maximum implosion speed of $7.1 \times 10^3 \textrm{ m/s}$, yielding $\textrm{M} = \dot{R}/c_{\infty} = 4.8$ relative to the ambient adiabatic speed of sound in water.
Additionally, the maximum radii ratio reached in the incompressible case is $R_m/R = 1460$, while in the compressible scenario, it is $R_m/R = 335$.
The details of the computations and numerical parameters used to obtain these results are discussed in Section~\ref{subsection_20_to_1_collapses}.
\par

\par

Figs.~\ref{fig:comparison_of_water} and \ref{fig:comparison_of_lithium} test the validity of our extension of the all-Mach solver to the compressibility asymptote.
The obtained solutions will aid in coupling the fluid dynamics to the molecular dynamics of the internal gas \cite{PhysRevLett.101.234301, PhysRevE.66.036310, BASS20082118, Schanz_2012}, which is part of future work.
When coupled with our results, these molecular dynamics simulations would help us understand the processes that determine how much energy density can be concentrated, as they would capture the motion and effects of internal dissipation within the gas.
One open question concerns the formation of a spherically imploding shock wave, which could further amplify energy focusing \cite{PhysRevLett.70.3424}.
The compressibility of the fluid slows down the collapse and may prevent shock formation.
However, fluids with different values of $B$ and $\Gamma$ could potentially allow for higher levels of energy density concentration.

%%%%%%%%%%%%%%%%%%%%%%%%%%%%%%%%%%%%%%%%%%%%%%%%%%%%%%%%%%%%%%%%%%%%%%%%%%%%%%%%%%%%%%%%%%%%%%%%%%%%%%%%%%%%%%%%%%%%%%%%%
% 
%                      THEORY
%
%%%%%%%%%%%%%%%%%%%%%%%%%%%%%%%%%%%%%%%%%%%%%%%%%%%%%%%%%%%%%%%%%%%%%%%%%%%%%%%%%%%%%%%%%%%%%%%%%%%%%%%%%%%%%%%%%%%%%%%%%

\section{Theory and the numerical method}
\subsection{Governing equations}

The spherically symmetric homentropic Euler equation of fluid mechanics, 
\begin{equation}
\label{Euler_equation}
\begin{gathered}
\rho \left( \frac{\partial v}{\partial t} + v\frac{\partial v}{\partial r}\right) = -\frac{\partial p}{\partial r},
\end{gathered}
\end{equation}
and the mass conservation law,
\begin{equation}
\label{mass_conservation}
\begin{gathered}
\frac{\partial \rho}{\partial t} + \frac{\partial (\rho v)}{\partial r} + \frac{2\rho v}{r} = 0,
\end{gathered}
\end{equation}
form the basis for our analysis \cite{Landau1987Fluid, Batchelor_2000}.
Here, $\rho(r,t), p(r,t),$ and $v(r,t)$ represent the mass density, pressure, and radial velocity at a given radial coordinate $r$ and time $t$.
These equations apply to the compressible fluid in the region outside the cavity, $r > R(t)$, referred to with subscript ``$f$'' in subsequent formulas, and to the gas inside the cavity, with $r<R(t)$, referred to with subscript ``$g$''.
Due to our assumption of homentropic flow, we do not consider the differential equation for specific entropy.
\par

To close the equations of motion, an expression for pressure is required.
For the compressible fluid in the region $r > R(t)$, the Tait-Murnaghan equation of state is assumed, as given in Eq.~\eqref{Tait_Murnaghan_EoS}.
Since the flow is homentropic, the functions $B$ and $\rho_0$ can be treated as constants.
For the gas inside the cavity, the assumed equation of state is that of an ideal adiabatic monatomic gas \cite{landau2013statistical, Batchelor_2000}, as given by:
\begin{equation}
\label{ideal_gas_EoS}
\begin{gathered}
p(\rho) = A \rho^{\gamma}, \quad
c^2 = \frac{\gamma p}{\rho}.
\end{gathered}
\end{equation}
In this case, the function $A$, which generally depends on specific entropy, can also be treated as a constant due to the homentropic nature of the flow, and $\gamma = 5/3$. 
Our choice for the gas is consistent with our goal of isolating the effects of compressibility while considering the simplest case where such effects can be observed.
\par

Although the general goal is to solve the full hydrodynamic Eqs.~\eqref{Euler_equation} and \eqref{mass_conservation} for the gas in the cavity, we will later, when discussing numerical results, introduce and explain the uniform bubble approximation used to avoid numerical artifacts and reduce computational time.
In this approach, the hydrodynamic equations for the gas inside the cavity are solved under the assumption that both the mass density and pressure are uniform throughout the entire cavity at all times, and consistent with the equation of state provided by Eq.~\eqref{ideal_gas_EoS}.
\par

\subsection{Boundary conditions}

The boundary conditions must be specified at the domain boundaries, as $r \to 0$ and as $r \to \infty$, as well as at the interface between the two fluids at $r=R(t)$.
As $r \to 0$, the boundary conditions are motivated by the requirement that the functions are smooth at the origin \cite{10.1121/1.3502464}, leading to $\partial \rho_g / \partial r = 0$ and $v_g = 0$.
As $r \to \infty$, the natural boundary conditions assume that the outer fluid is at rest at ambient pressure, but instead, we use $\partial \rho_f / \partial r = 0$ and $\partial v_f/\partial r = 0$, which can be derived from the rest condition by assuming smoothness of the solutions.
These latter boundary conditions exhibit better numerical properties when applied to a large but finite computational domain.
Furthermore, varying the size of the numerical domain has been found not to affect the obtained solutions, validating our choice of boundary conditions and domain size.
Finally, at the interface, as we ignore viscosity, mass transfer, and surface tension effects, the boundary conditions are \cite{Batchelor_2000} $v_g = v_f$ and $p_g(\rho_g) = p_f(\rho_f)$.
Because the radial velocity is continuous across the interface, the evolution of the cavity wall can be described using the velocity fields of either fluid:
\begin{equation}
\label{velocity_bc}
\begin{gathered}
\dot{R}(t) = v_g(R(t), t) = v_f(R(t), t).
\end{gathered}
\end{equation}
\par

\subsection{Initial conditions}

The initial conditions must also be provided for the hydrodynamic functions of both fluids and for the initial state of the cavity wall.
Our choice follows the approach used in Ref.~\onlinecite{10.1121/1.3502464}.
We take the initial state to be at a time $t=0$, where the cavity is at its maximum radius $R_m$, and the collapse starts from rest.
In that case, $R(t=0) = R_m$, $\dot{R}(t=0) = 0$, $v_g(r, t=0) = 0$ for all $r < R(t=0)$, and $v_f(r, t=0) = 0$ for all $r > R(t=0)$.
The initial conditions for the pressure are estimated by assuming that the motion near the beginning of the collapse is incompressible, as everything starts from rest.
For the outside fluid in the region $r > R(t)$, the incompressible solution at all times is given by \cite{doi:10.1080/14786440808635681, Batchelor_2000}
\begin{equation}
\label{velocity_incompressible}
\begin{gathered}
v_f(r,t) = \dot{R}(t) \left( \frac{R(t)}{r} \right)^2,
\end{gathered}
\end{equation}
\begin{equation}
\label{pressure_incompressible}
\begin{gathered}
p_f(r,t) = \frac{\rho_{\infty}\dot{R}^2(t)}{2} \left( \frac{R(t)}{r} - \left(\frac{R(t)}{r}\right)^4 \right) + p_{\infty} \left( 1 - \frac{R(t)}{r} \right) + p_g(R(t)) \frac{R(t)}{r},
\end{gathered}
\end{equation}
\begin{equation}
\label{radius_incompressible}
\begin{gathered}
\ddot{R}(t) R(t) + \frac{3}{2} \dot{R}^2(t) = - \frac{\left( p_{\infty} - p_g(R(t)) \right)}{\rho_{\infty}}.
\end{gathered}
\end{equation}
\par

In these expressions, $p_g(R(t))$ represents the uniform pressure for the gas in the bubble at any given radius $R(t)$.
If the mass density of the gas $\rho_g(R(t))$ is also assumed to be uniform, then one can use the mass conservation law to relate it to the ambient parameters, where the ambient radius of the bubble is $R_0$, and the ambient mass density of the gas is $\rho_{g,0}$.
The ambient mass density of the gas can be related to the ambient temperature $T_{\infty}$ using the ideal gas law, where $k_B$ is the Boltzmann constant and $m$ is the mass of a single gas atom:
\begin{equation}
\label{density_incompressible}
\begin{gathered}
\rho_g(R(t)) = \rho_{g,0} \left( \frac{R_0}{R(t)} \right)^3, \quad \rho_{g, 0} = \frac{p_{\infty} m}{k_B T_{\infty}}.
\end{gathered}
\end{equation}
Together with the assumed equation of state in Eq.~\eqref{ideal_gas_EoS}, this provides the uniform pressure for the ideal gas under consideration:
\begin{equation}
\label{pressure_gas_incompressible}
\begin{gathered}
p_g(R(t)) = p_{\infty} \left( \frac{\rho_g(R(t))}{\rho_{g,0}} \right)^{\gamma} = p_{\infty} \left( \frac{R_0}{R(t)} \right)^{3\gamma}.
\end{gathered}
\end{equation}
\par

Using these results, we assume the initial mass density and pressure profiles for the gas to be uniform for all $r < R(t=0)$, with values given by Eqs.~\eqref{density_incompressible} and \eqref{pressure_gas_incompressible}, respectively, when evaluated at $R(t = 0)$.
The initial pressure profile for the outside fluid for all $r > R(t=0)$ is then derived from Eq.~\eqref{pressure_incompressible}, using the initial conditions of the cavity wall along with Eq.~\eqref{pressure_gas_incompressible}:
\begin{equation}
\label{initial_pressure_profile}
\begin{gathered}
p_f(r,t = 0) = p_{\infty} + \frac{ p_{\infty} R_m}{r} \left( \left( \frac{R_0}{R_m} \right)^{3\gamma} - 1 \right).
\end{gathered}
\end{equation}
From here, the initial mass density profile for the outside fluid, $\rho_f(r, t=0)$, can be obtained by inverting Eq.~\eqref{initial_pressure_profile} using the equation of state provided in Eq.~\eqref{Tait_Murnaghan_EoS}.
\par

Our interest in cavity collapse is driven by our desire to understand the limits of energy focusing achievable through sonoluminescence.
As mentioned, this requires a parameter space where $R_m/R_0 > 8$.
In this case, at least for water, Eqs.~\eqref{velocity_incompressible}, \eqref{pressure_incompressible}, and \eqref{radius_incompressible} become incorrect at long times.
Nevertheless, comparing our calculations to these incompressible solutions helps us understand when the effects of compressibility become important and how significant the differences are. 
Additionally, we will later discuss an improvement to the initialization of our computation by starting from a radius lower than $R_m$, as the effects of compressibility are negligible during the first stage of the collapse.
In that case, the initial conditions will be modified and estimated from the incompressible solutions, allowing for faster numerical calculations.
\par

\subsection{Numerical method}

To numerically solve the proposed problem, we use an all-Mach formulation for simulating bubble dynamics that employs the volume-of-fluid method \cite{FUSTER2018752} and is implemented in the Basilisk framework \cite{POPINET2015336}. 
We modify this formulation for the case of spherical symmetry in the same way as described in Ref.~\onlinecite{SAADE2023111865}.
We now summarize the key ideas of this formulation.
\par

A one-fluid formulation for handling hydrodynamic quantities across the entire domain can be obtained by introducing the Heaviside function, $\mathcal{H}(r,t)$, which takes the value 0 in the gas region and 1 in the surrounding fluid \cite{FUSTER2018752}. 
In this approach, the interface between the gas and fluid is represented as a discontinuity in the Heaviside function. 
The average $\overline{\phi}$ of a quantity is defined as $\overline{\phi}=(1-\mathcal{H})\phi_g + \mathcal{H} \phi_f$, where $\phi_g$ and $\phi_f$ are the corresponding quantities in the gas and fluid, respectively. 
This formulation provides a unified treatment across the domain, avoiding the need to solve separate governing equations in the gas and liquid regions. \par

In the discrete setting, a finite volume discretization is employed \cite{FUSTER2018752}.
As part of this, a time-dependent color function, $C$, corresponding to the Heaviside function, $\mathcal{H}$, is introduced.
The color function $C$ represents the volume fraction of the surrounding fluid phase within each grid cell relative to the total cell volume.
Similar to the Heaviside function, it takes the value $0$ in a cell fully within the bubble and $1$ in a cell entirely in the surrounding fluid.
Only in cells that are mixed --- i.e., containing the bubble wall --- does $C$ take on values strictly between 0 and 1.
By leveraging volume fractions, we can resolve the location of the bubble wall at the subgrid level, improving spatial accuracy in tracking the interface.
The evolution of the color function is the same as for the Heaviside function and is governed by an advection equation:
\begin{equation}
\label{color_function_advection}
\begin{gathered}
\frac{\partial C}{\partial t} + \frac{\partial \left( C\overline{v} \right)}{\partial r} = C \frac{\partial \overline{v}}{\partial r},
\end{gathered}
\end{equation}
where $\overline{v}$ is the averaged velocity field.
As before for mass density,  the boundary conditions are specified by assuming the smoothness of solutions and with the goal of improving numerical properties, leading to $\partial C/\partial r = 0$ as $r \to 0$ and as $r \to \infty$.
The initial condition for a given uniform spatial grid with intervals of width $\Delta r$ is defined as $C(r, t=0) = 0$ for $r < \left( R(t=0) - \Delta r/2 \right)$, $C(r, t=0) = 1$ for $r > \left( R(t=0) + \Delta r/2 \right)$, and $C(r, t=0)$ is linearly interpolated between $0$ and $1$ in the region $\left( R(t=0) - \Delta r/2 \right) \leq r \leq \left( R(t=0) + \Delta r/2 \right)$ to ensure continuity. 
Finally, the interface between the two fluids, represented by the cavity radius $R(t)$ at time $t$, can be numerically evaluated as follows:
\begin{equation}
\label{numerically_finding_R_t}
\begin{gathered}
R(t) = \Delta r \sum \limits_{i} (1 - C_i(t)),
\end{gathered}
\end{equation}
where $i$ represents a grid interval in the numerical domain, and $C_i(t)$ is the value of the color function in that interval at time $t$.
\par

For a homentropic flow, the Euler equation and the mass conservation law, given in Eqs.~\eqref{Euler_equation} and \eqref{mass_conservation}, respectively, completely specify the problem, provided that the equation of state for the pressure is known.
Nevertheless, in addition to solving Eq.~\eqref{color_function_advection} and the averaged versions of Eqs.~\eqref{Euler_equation} and \eqref{mass_conservation}, the energy conservation law is also numerically solved in its averaged form.
Basilisk includes a solution of the energy equation to support thermal conduction, although this feature is not used in the present work.
\par 

As previously mentioned, a uniform spatial grid with grid interval width $\Delta r$ is employed.
Future results could be improved by implementing adaptive gridding.
The computational domain is finite, defined as $0 < r < \lambda \times R(t=0)$, where $\lambda$ is a numerical parameter that specifies the cutoff for sufficiently large values of $r$.
The initial radius in our simulations is typically $R(t=0) = R_m$ or, as will be discussed later, $R(t=0) = R_m/5$.
The boundary conditions, which theoretically apply in the limit $r \to \infty$, are applied at the cutoff $r = \lambda \times R(t=0)$ in our formulation \cite{10.1121/1.3502464, SAADE2023111865}.
\par

For quick calculations, it is advantageous to choose $\lambda$ as small as possible without introducing significant errors from truncating the semi-infinite domain to a finite one.
In the result shown in Fig.~\ref{fig:comparison_of_water} for water, $\lambda = 8$ was used, while in Fig.~\ref{fig:comparison_of_lithium} for liquid lithium, $\lambda = 16$ was employed.
These values were determined through numerical convergence studies, which are discussed below.
For our specific problems, and given these values of $\lambda$, the errors resulting from domain truncation are negligible compared to the spatial discretization errors.
We note that liquid lithium is anticipated to require a larger computational domain due to its higher speed of sound.
\par

A time step size $\Delta t$ during the computation is not fixed and is determined by combining a fixed time step size parameter with adaptive time-stepping schemes.
The time step size parameter $\Delta t_{\textrm{max}}$ is fixed throughout the calculation and determines the largest permitted time step size for a given calculation.
Two adaptive time step size schemes are used, both inspired by the CFL condition \cite{anderson1995computational}.
In one scheme, the time step size at each iteration is estimated as $\Delta t_{\overline{v}} = (1/2) \Delta r / |\overline{v}|_{\textrm{max}}$, where $|\overline{v}|_{\textrm{max}}$ is the largest magnitude of the average radial velocity across all grid intervals.
The other scheme uses $\Delta t_{\overline{c}} = (1/2) \Delta r / |\overline{c}|_{\textrm{max}}$, where $|\overline{c}|_{\textrm{max}}$ is the largest magnitude of the average speed of sound across all grid intervals.
At each instant, the time step size $\Delta t$ is determined as the smallest value among $\Delta t_{\textrm{max}}, \Delta t_{\overline{v}}$, and $\Delta t_{\overline{c}}$.
The adaptive methods provide simple estimates for the time step size suitable for the complicated nonlinear problem at hand, while $\Delta t_{\textrm{max}}$ allows for further refinement until the desired level of accuracy is achieved.
Additionally, a small value of $\Delta t_{\textrm{max}}$ leads to fewer iterations and better convergence of the underlying multigrid Poisson solver.
\par

To determine the optimal value for $\lambda$, we begin with $\lambda = 4$ and perform runs by varying $\Delta r$ without incorporating $\Delta t_{\text{max}}$ in the calculation of $\Delta t$.
We identify the smallest $\Delta r$ that still allows a rapid solution.
Next, we set $\Delta t_{\textrm{max}}$ to the smallest $\Delta t$ obtained from a run using only adaptive time step size schemes.
Following this, we analyze numerical convergence in $\lambda$ by fixing the identified $\Delta r$ and $\Delta t_{\text{max}}$, and increasing $\lambda$ by factors of two.
Convergence is evaluated using four parameters: the smallest achieved radius, $R_{\text{min}}$, and the time at which it is achieved, $t_{\text{min}}$, as well as the largest velocity during a collapse, $\dot{R}_{\text{max}}$, and the time at which it is achieved, $t_{\dot{R}_\textrm{max}}$.
The velocity curves $\dot{R}(t)$ are computed using second-order accurate central differences. 
For an example of such a convergence study, refer to the discussion for Fig.~\ref{fig:water_convergence_in_lambda}.
\par

After the optimal value for $\lambda$ is determined and fixed, we perform numerical convergence studies on $\Delta r$ and $\Delta t_{\text{max}}$.
The initial values of these parameters are chosen as in the $\lambda$ convergence studies. 
Subsequently, both $\Delta r$ and $\Delta t_{\text{max}}$ are decreased simultaneously by factors of two until the desired accuracy is achieved or until the limits of computational resources are reached. 
Simultaneous refinement saves time spent on studying the convergence of the numerical solutions and has also been observed to enhance the convergence for the $\dot{R}(t)$ curves. 
To assess convergence, we study the same four parameters: $R_{\textrm{min}}$, $t_{\textrm{min}}$, $\dot{R}_{\textrm{max}}$, and $t_{\dot{R}_\textrm{max}}$. 
For examples, see the discussion for Figs.~\ref{fig:water_convergence_in_dr} and \ref{fig:water_convergence_in_dr_velocity}.
\par

The numerical simulations were performed on a 13th Gen Intel Core i9-13900 2.00 GHz processor workstation with 24 cores.
The Basilisk framework supports parallel computation.
We took advantage of this capability and performed all computations with 4 threads.
Using more threads did not provide a sufficient speedup to justify the additional resource usage.

%%%%%%%%%%%%%%%%%%%%%%%%%%%%%%%%%%%%%%%%%%%%%%%%%%%%%%%%%%%%%%%%%%%%%%%%%%%%%%%%%%%%%%%%%%%%%%%%%%%%%%%%%%%%%%%%%%%%%%%%%
% 
%                      NUMERICAL RESULTS
%
%%%%%%%%%%%%%%%%%%%%%%%%%%%%%%%%%%%%%%%%%%%%%%%%%%%%%%%%%%%%%%%%%%%%%%%%%%%%%%%%%%%%%%%%%%%%%%%%%%%%%%%%%%%%%%%%%%%%%%%%%

\section{Numerical Results for a Gas Filled Cavity}
Our goal is to describe collapses that are sufficiently strong for their motion to resemble that of light-emitting bubbles. 
To this end, we calculate cases where $R_m/R_0 = 10$ and $20$.
Even in the case where the expansion ratio is $10$, and even at early times when $R(t)$ is close to $R_m$ so that the speed of the bubble wall is very small, we encounter an unphysical response when solving the full hydrodynamic equations that include Eqs.~\eqref{Euler_equation} and \eqref{mass_conservation} for the gas inside the bubble.
To obtain physically meaningful solutions that are consistent with the asymptotic regime approached at small $R$, we introduce the uniform bubble approximation for the interior gas.
This method is then applied to a stronger collapse where $R_m/R_0 = 20$, showing excellent numerical properties.

\subsection{Unphysical solver output for collapses with $R_m/R_0 = 10$ when solving the full hydrodynamic equations inside the cavity}

Before considering a more rapid collapse where $R_m/R_0 = 20$, we first solved a smaller but still significant collapse with $R_m/R_0 = 10$.
The gas inside the cavity is assumed to be xenon, while the outside fluid is taken to be water.
The full hydrodynamic equations are solved numerically for both the gas and the fluid outside the cavity, using the previously described boundary conditions.
The initial conditions correspond to both the gas and fluid starting from rest, with the cavity at its maximum radius $R_m$.
The ambient physical parameters, $R_0 = 2.2\textrm{ }\upmu\textrm{m}$, $p_{\infty} = 1\textrm{ atm}$, and $T_{\infty} = 293\textrm{ K}$, are motivated by values found in experiments \cite{PhysRevLett.101.234301}.
\par

We start by exploring the results of the full all-Mach solver for a coarsely discretized run where $\lambda = 4$, $\Delta r = R_m/2^{11}=10R_0/2^{11}$, and $\Delta t_{\textrm{max}}$ is not used in computing the time step.
With these parameters, the code aborts at $t_* = 0.922 R_0 \sqrt{\rho_{\infty}/p_{\infty}}$, well before reaching the expected minimum radius during the collapse.
Specifically, during this initial stage of the collapse, where $R(t)$ is still close to $R_m$, the mass density profile of the gas quickly transitions from uniform to developing a region of extremely low density near the cavity wall, while in the same region, the computed speed of sound becomes extremely high.
As a result, the adaptive time step size, based on the speed of sound, becomes excessively small, making the computation infeasible and causing it to abort.
\par

To avoid the issue of the extremely small time step size, we explored modifying the adaptive time step scheme $\Delta t_{\overline{c}}$ to use the ambient speed of sound of the surrounding fluid, given by Eq.~\eqref{Tait_Murnaghan_speed_of_sound}, instead of $|\overline{c}|_{\textrm{max}}$.
This is motivated by the fact that the speed of sound in the gas is much lower in the initial stages of the collapse than the ambient speed of sound of the outside fluid.
This choice also makes $\Delta t_{\overline{c}}$ constant during the calculation, allowing computations to extend beyond $t_*$ and enabling an analysis of whether the observed issue can be resolved with more refined calculations.
\par

Performing the run with the same numerical parameters but with the modified adaptive time step scheme $\Delta t_{\overline{c}}$ allows calculations to proceed further than before.
However, even with this adjustment, the simulation aborts at $t_* = 1.51 R_0 \sqrt{\rho_{\infty}/p_{\infty}}$, which is well before the expected minimum radius.
Now the issue arises because the solver for the pressure values fails to achieve sufficient convergence.
Moreover, the near-vacuum region in the gas near the bubble wall persists.
We now examine the convergence of the method, with particular attention paid to the observed vanishing density.
Specifically, we wish to determine whether we can resolve the issue by increasing the resolution of our numerical solution by considering larger values of $\lambda$ and smaller values of $\Delta t_{\textrm{max}}$ and $\Delta r$.
As we will discuss next, achieving sufficient resolution to remove the observed numerical artifact is not feasible within our computational budget.
\par

As discussed earlier, to determine a suitable value for $\Delta t_{\textrm{max}}$, we avoid relying on the computed speed of sound in the gas due to the blowup phenomenon near the bubble wall.
Since the speed of sound in the liquid is initially greater than that in the gas and remains so for much of the collapse, we take $\Delta t_{\textrm{max}} = (1/2) \Delta r / c_{\infty} = 1.65 \times 10^{-5} R_0 \sqrt{\rho_{\infty}/p_{\infty}}$ at the coarsest grid resolution.
This value is then decreased in powers of two, leading to $\Delta t_{\textrm{max}} = 1.65/16 \times 10^{-5} R_0 \sqrt{\rho_{\infty}/p_{\infty}}$ at the most refined resolution.
The refinement changes the time at which the solver fails, so to ensure comparability, we evaluate all runs until $t = 1.50 R_0 \sqrt{\rho_{\infty}/p_{\infty}}$, which is near $t_*$ of the least refined run.
All simulations successfully reach this time.
Although the curves for $R(t)$ show negligible change across refinements, with $R_{\textrm{min}}$ and $t_{\textrm{min}}$ varying by less than $10^{-4}\%$, $\dot{R}_{\textrm{max}}$ and $t_{\dot{R}_\textrm{max}}$ differ by over $10\%$ even between the two most refined runs.
Crucially, the numerical artifacts in the hydrodynamic profiles, described previously, persist without improvement.
\par

We also consider the refinement of $\lambda$ with the fixed value $\Delta t_{\textrm{max}}= 1.65 \times 10^{-5} R_0 \sqrt{\rho_{\infty}/p_{\infty}}$.
As the velocity values are unreliable due to the developed numerical artifacts, we only use $R_{\textrm{min}}$ and $t_{\textrm{min}}$ as the numerical parameters for testing convergence.
The value of $\lambda = 4$ is increased by factors of two until $\lambda = 32$ is achieved, and we observe that this does not change the time at which simulation aborts, $t_* = 1.51 R_0 \sqrt{\rho_{\infty}/p_{\infty}}$.
Considering all runs until the time $t = 1.50 R_0 \sqrt{\rho_{\infty}/p_{\infty}}$, we find that with each level of refinement, $R_{\textrm{min}}$ and $t_{\textrm{min}}$ change by less than $10^{-3}\%$, while the numerical artifacts found in the hydrodynamic profiles again do not improve.
\par

\begin{figure}
\includegraphics[width=5.2in, height=4.8in]{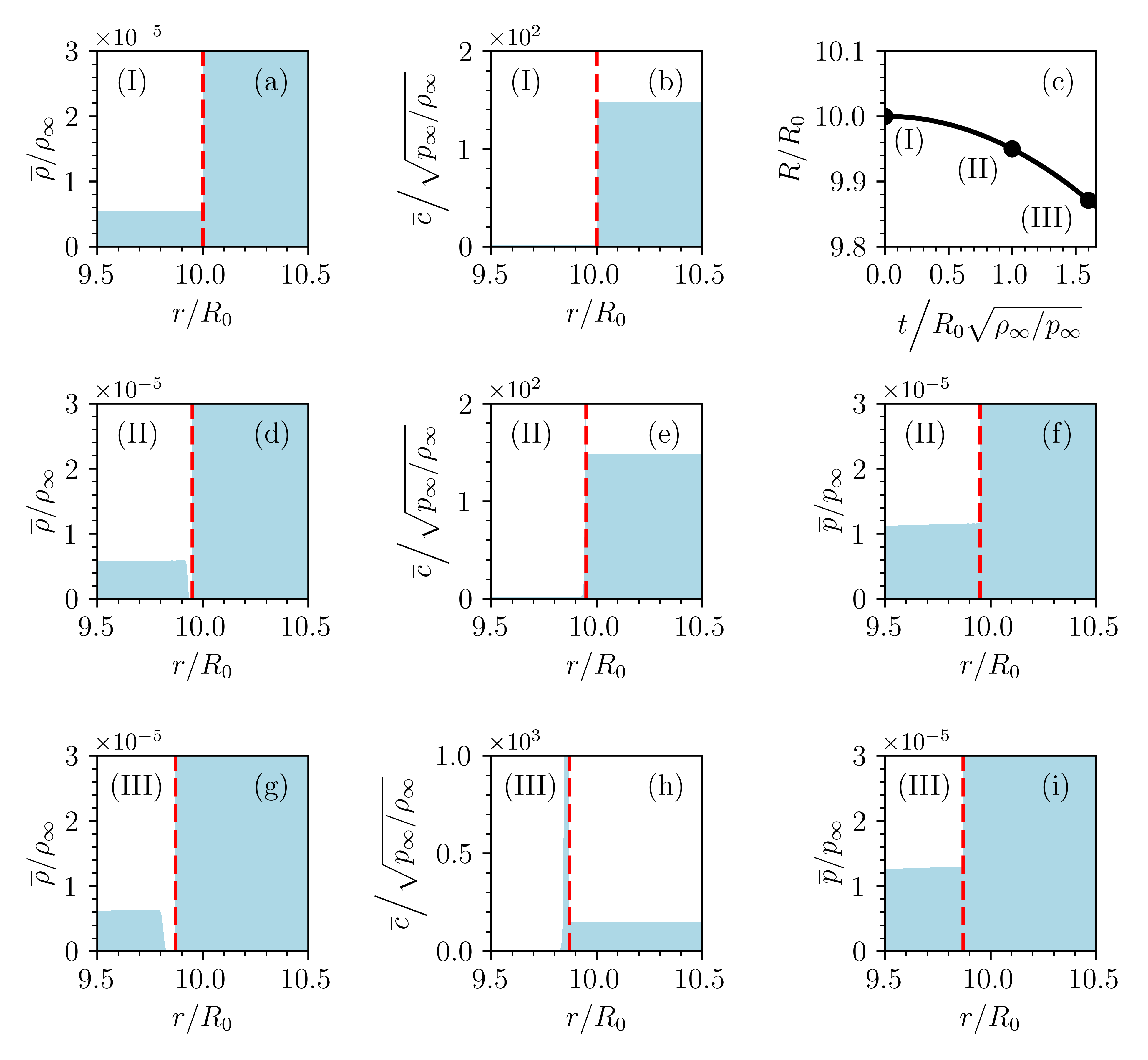} %for two-column figure, 6.69 inches x 8.25 inches. 
\caption{
Implosion of an ideal gas xenon bubble surrounded by compressible water with $R_m/R_0 = 10$ and ambient parameters $R_0 = 2.2\textrm{ }\upmu\textrm{m}$, $p_{\infty} = 1\textrm{ atm}$, and $T_{\infty} = 293\textrm{ K}$.
The full hydrodynamic equations given in Eqs.~\eqref{Euler_equation} and \eqref{mass_conservation} are solved for both the gas and the surrounding fluid, starting from rest at the maximum radius $R_m$.
The numerical parameters are $\lambda = 4$, $\Delta r = R_m/2^{15} = 10R_0/2^{15}$, and $\Delta t_{\textrm{max}} = 1.0 \times 10^{-6}$ $R_0 \sqrt{\rho_{\infty}/p_{\infty}}$.
The figure shows the solution at three times $t = 0.0, 1.0, 1.6$, in units of $R_0 \sqrt{\rho_{\infty}/p_{\infty}}$, corresponding to (I), (II), (III), respectively.
Figures (a), (d), (g) show profiles of the average mass density, $\overline{\rho}$; (b), (e), (h) show the average speed of sound, $\overline{c}$; and (f) and (i) show the average pressure, $\overline{p}$, with the red dashed line indicating the location of the cavity wall at each time.
In (c), $R(t)$ is shown.
As observed, shortly after the collapse is initiated, while $R(t)$ remains close to $R_m$, a region of low mass density and high speed of sound develops in the gas near the wall.
}
\label{fig:basilisk_from_Rmax}
\end{figure}

Finally, convergence with respect to the simultaneous refinement of $\Delta r$ and $\Delta t_{\textrm{max}}$ was considered, starting with values $\lambda = 4$, $\Delta r = R_m/2^{11}=10R_0/2^{11}$, and $\Delta t_{\textrm{max}} = 1.65 \times 10^{-5} R_0 \sqrt{\rho_{\infty}/p_{\infty}}$, which required 42 minutes of computational time.
The most refined calculation, with $\Delta r = R_m/2^{15}=10R_0/2^{15}$ and $\Delta t_{\textrm{max}} = 1.65/16 \times 10^{-5} R_0 \sqrt{\rho_{\infty}/p_{\infty}}$, took 3 days to complete.
The time at which the pressure solver fails to converge varies with refinement, sometimes occurring earlier with more refined grids.
In these runs, the earliest abort time is $t_* = 1.41 R_0 \sqrt{\rho_{\infty}/p_{\infty}}$, so the comparison is made up to $t = 1.40 R_0 \sqrt{\rho_{\infty}/p_{\infty}}$.
With each level of refinement, the $R(t)$ curves exhibit minimal change: $R_{\textrm{min}}$ and $t_{\textrm{min}}$ vary by less than $10^{-5}\%$, although the hydrodynamic profiles show significant improvement.
The results for mass density, speed of sound, and pressure, along with the cavity motion radius curve from the most refined computation, are shown in Fig.~\ref{fig:basilisk_from_Rmax}.
Comparing the data at $t = 1.50 R_0 \sqrt{\rho_{\infty}/p_{\infty}}$ for the least and most refined runs, we observe that the size of the anomalous region of low density and high speed of sound near the bubble wall decreases by a factor of $10$, from $0.5R_0$ to $0.05R_0$. 
Nevertheless, even the most refined calculation still does not allow for the simulation of the collapse beyond its initial stages.

\par

\subsection{The uniform bubble approximation for collapses with $R_m/R_0 = 10$}

It is known that a piston moving into a gas at subsonic speed does not create a vacuum \cite{Landau1987Fluid}.
Despite our best efforts, we were unable to apply sufficient refinement to eliminate this spurious behavior in the sonoluminescence limit, leading us to conclude that this approach is unlikely to yield productive results.
Therefore, we seek an approximation for the interior gas that allows for the simulation of compressible fluid motion during strong collapses. 
\par

Motivated by the low Mach number expansion, which we take to be accurate near the maximum radius \cite{10.1063/1.858700}, we assume that the mass density and pressure inside the gas are uniform but time-dependent at all times.
Although the cavity’s velocity increases by many orders of magnitude at later times, leading to potential failure of this approximation, it allows us to consider the effects of compressibility of the outside fluid on the collapse while bypassing the previously mentioned numerical artifacts observed in the computed mass density and the speed of sound.
This approach is similar to the incompressible solution given in Eqs.~\eqref{pressure_incompressible} and \eqref{radius_incompressible}, where the pressure in the gas is also assumed to be uniform.
A weakness of such a method is not capturing the effects of nonuniformities in the gas, such as the formation of shock waves.
To resolve such phenomena, the full problem must be solved.
\par

To numerically implement the uniform bubble approximation, we begin each time step by allowing the original all-Mach solver to execute all its procedures, including solving Eq.~\eqref{color_function_advection} and the averaged Eqs.~\eqref{Euler_equation} and \eqref{mass_conservation} for all $r > 0$, which includes the region containing the gas.
These procedures also update the bubble's location, $R(t)$, and its velocity, $\dot{R}(t)$.
The value of $R(t)$ is numerically determined using Eq.~\eqref{numerically_finding_R_t}, while $\dot{R}(t)$ is estimated by identifying the grid cell $i$ with radial coordinates $r_i < r < r_{i+1} = r_i + \Delta r$ that contains $R(t)$, and then using this cell's grid-centered average radial velocity $\overline{v}_i$ to estimate $\dot{R}(t)$.
Before completing the time step, we also identify all grid cells $i$ where $C_i = 0$, indicating that these cells are occupied solely by the gas.
In all such grid cells, in accordance with our uniform bubble approximation, we set the values of mass density according to Eq.~\eqref{density_incompressible} and pressure according to Eq.~\eqref{pressure_gas_incompressible}, both evaluated at the current radius $R(t)$.
Additionally, the velocity field in the gas is overwritten based on the low Mach number result \cite{10.1063/1.858700}
\begin{equation}
\label{low_mach_velocity_in_the_gas}
v_g(r,t) = \dot{R}(t) \frac{r}{R(t)}.
\end{equation}
\par

\begin{figure}
\includegraphics[width=5.2in, height=4.8in]{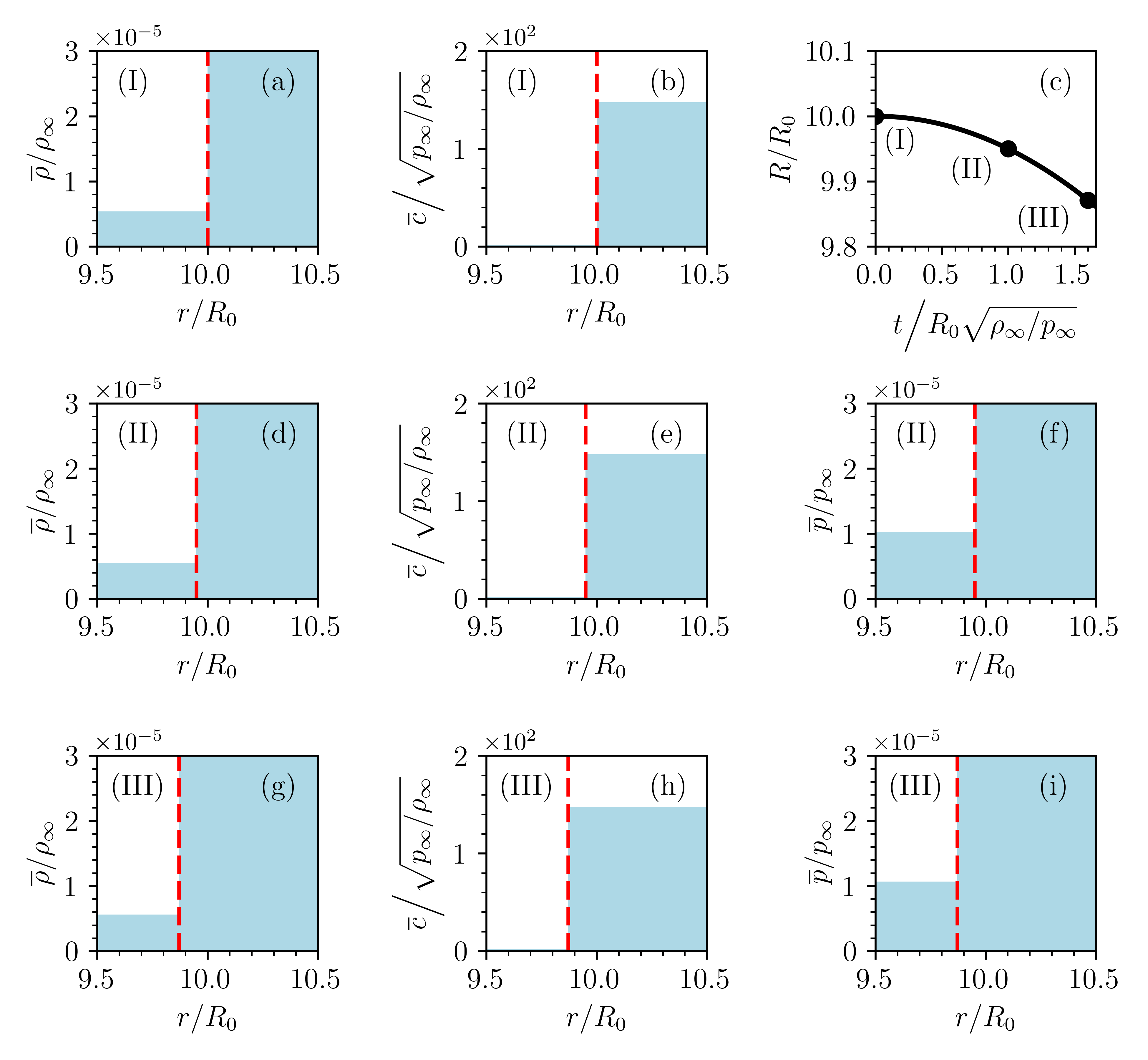} %for two-column figure, 6.69 inches x 8.25 inches. 
\caption{
Implosion of an ideal gas xenon bubble surrounded by compressible water with $R_m/R_0 = 10$ and ambient parameters $R_0 = 2.2\textrm{ }\upmu\textrm{m}$, $p_{\infty} = 1\textrm{ atm}$, and $T_{\infty} = 293\textrm{ K}$.
The uniform bubble approximation for the motion of the gas is assumed, and the motion starts from rest at the maximum radius $R_m$.
The numerical parameters are $\lambda = 4$, $\Delta r = R_m/2^{15} = 10R_0/2^{15}$, and $\Delta t_{\textrm{max}} = 1.0 \times 10^{-6}$ $R_0 \sqrt{\rho_{\infty}/p_{\infty}}$.
The figure shows the solution at three times $t = 0.0, 1.0, 1.6$, in units of $R_0 \sqrt{\rho_{\infty}/p_{\infty}}$, corresponding to (I), (II), (III), respectively. 
Figures (a), (d), (g) show profiles of the average mass density, $\overline{\rho}$; (b), (e), (h) show the average speed of sound, $\overline{c}$; and (f) and (i) show the average pressure, $\overline{p}$, with the red dashed line indicating the location of the cavity wall at each time. 
In (c), $R(t)$ is shown.
}
\label{fig:uniform_from_Rmax}
\end{figure}

To see how such an approximation influences the numerical solutions, we consider the same problem with the same physical and numerical parameters as the one considered previously, which was solved using the original all-Mach solver formulation.
We continue to use the modified adaptive time step scheme $\Delta t_{\overline{c}}$.
The results can be seen in Fig.~\ref{fig:uniform_from_Rmax} where we observe that the low mass density and high speed of sound region no longer appears.
Previously, the computation with this resolution aborted at $t_* = 1.66 R_0 \sqrt{\rho_{\infty}/p_{\infty}}$, so we performed the modified computation until the same time.
The new calculation took 5 hours to complete, which represents an improvement of approximately a factor of $15$ in the computational time.
As will be discussed in detail next, with this approximation it becomes possible to perform computations up to the minimum of the $R(t)$ curve and capture the long-time behavior, such as the outgoing waves in the compressible fluid, that happen after the minimum radius.

\par

\subsection{The uniform bubble approximation for collapses with $R_m/R_0 = 20$}
\label{subsection_20_to_1_collapses}

In our simulations using the uniform bubble approximation, we observed that during the initial stage of the collapse, even with compressible effects present in the equations, the behavior closely follows that of the incompressible collapse.
This motivated us to introduce an additional approximation by starting the bubble from a smaller radius.
Specifically, we set for our calculations $R(t = 0) = R_m/5$.
Consequently, the initial conditions, which originally correspond to the cavity starting at the maximum radius and from rest, need to be modified.
\par

For the cavity wall, we take $R(t=0)=R_m/5$ and $\dot{R}(t=0)$ from the incompressible result \cite{doi:10.1080/14786440808635681}
\begin{equation}
\dot{R}(t) = -\sqrt{\frac{2p_{\infty}}{3\rho_{\infty}}} \sqrt{ \left( \frac{R_m}{R(t)} \right)^3 - 1}, 
\end{equation}
which assumes that the pressure in the gas is negligible compared to $p_{\infty}$.
This assumption is justified for strong collapses, as indicated by Eq.~\eqref{pressure_gas_incompressible}.
\par

Using the uniform bubble approximation, the initial values for the uniform mass density and pressure profiles of the gas in the cavity are given by Eqs.~\eqref{density_incompressible} and \eqref{pressure_gas_incompressible}, while the velocity profile is given by the low Mach solution in Eq.~\eqref{low_mach_velocity_in_the_gas}.
Since the motion of the outside fluid remains incompressible, the velocity and pressure profiles are determined by the incompressible solutions in Eqs.~\eqref{velocity_incompressible} and \eqref{pressure_incompressible}, while the mass density profile is obtained by inverting the initial pressure profile using the equation of state given in Eq.~\eqref{Tait_Murnaghan_EoS}.
Finally, as the uniform bubble approximation resolves the issue with the speed of sound, we revert to the original adaptive time step size scheme $\Delta t_{\overline{c}} = (1/2) \Delta r / |\overline{c}|_{\textrm{max}}$.
\par

\begin{figure}
\includegraphics[width=4.8in, height=2.0in]{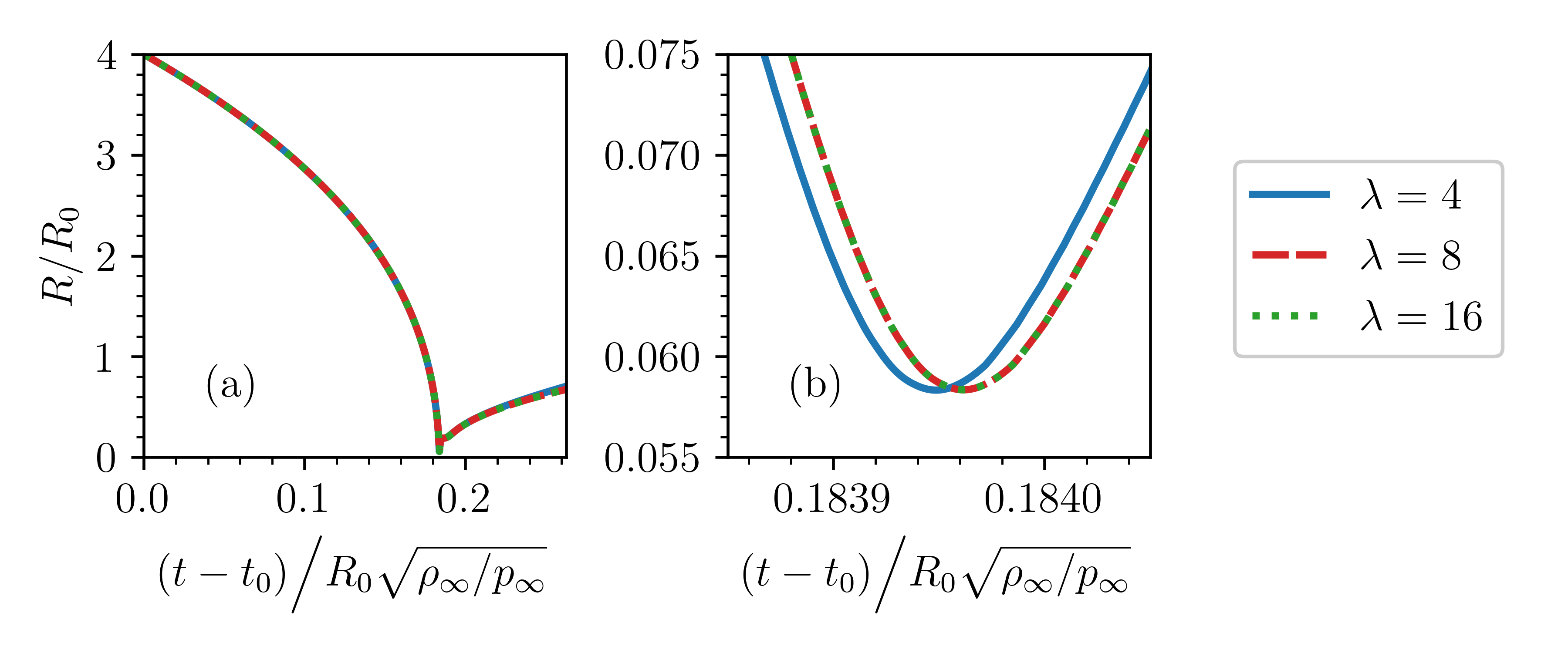} %for two-column figure, 6.69 inches x 8.25 inches. 
\caption{
The dependence of the strong collapse on the size of the numerical domain, parametrized by $\lambda$, for a case with $R_m/R_0 = 20$ involving an ideal xenon gas cavity surrounded by compressible water.
The uniform bubble approximation for the gas motion is assumed, with the motion starting from $R(t_0) = R_m/5$.
The ambient parameters are $R_0 = 2.2\textrm{ }\upmu\textrm{m}$, $p_{\infty} = 1\textrm{ atm}$, and $T_{\infty} = 293\textrm{ K}$.
The numerical parameters are $\Delta r = R_0/2^{9}=R_0/512$ and $\Delta t_{\textrm{max}} = 5.9 \times 10^{-7}$ $R_0 \sqrt{\rho_{\infty}/p_{\infty}}$.
In (b), the results shown in (a) are zoomed in near the minimum radius of the collapse.
}
\label{fig:water_convergence_in_lambda}
\end{figure}

Using the two introduced approximations: uniform gas in the bubble and starting the calculation from $R_m/5$, we can obtain results for stronger collapses, where $R_m/R_0 = 20$.
First, we consider the case of an ideal xenon gas bubble surrounded by water.
The ambient parameters are the same as before, with $R_0 = 2.2\textrm{ }\upmu\textrm{m}$, $p_{\infty} = 1\textrm{ atm}$, and $T_{\infty} = 293\textrm{ K}$.
The convergence with respect to $\lambda$ was performed with $\Delta r = R_0/2^{9}=R_0/512$ and $\Delta t_{\textrm{max}} = 5.9 \times 10^{-7}$ $R_0 \sqrt{\rho_{\infty}/p_{\infty}}$.
For $\lambda = 4$, the computation took around 3 minutes, while the largest value, $\lambda = 16$, ran for approximately one hour.
It was computed that all four parameters, $R_{\textrm{min}}$, $t_{\textrm{min}}$, $\dot{R}_{\textrm{max}}$, and $t_{\dot{R}_\textrm{max}}$, experience negligible change when increasing $\lambda$ from $8$ to $16$.
Therefore, $\lambda = 8$ is used for the following calculations.
The results for $R(t)$ can be seen in Fig.~\ref{fig:water_convergence_in_lambda}.

\begin{figure}
\includegraphics[width=4.8in, height=2.0in]{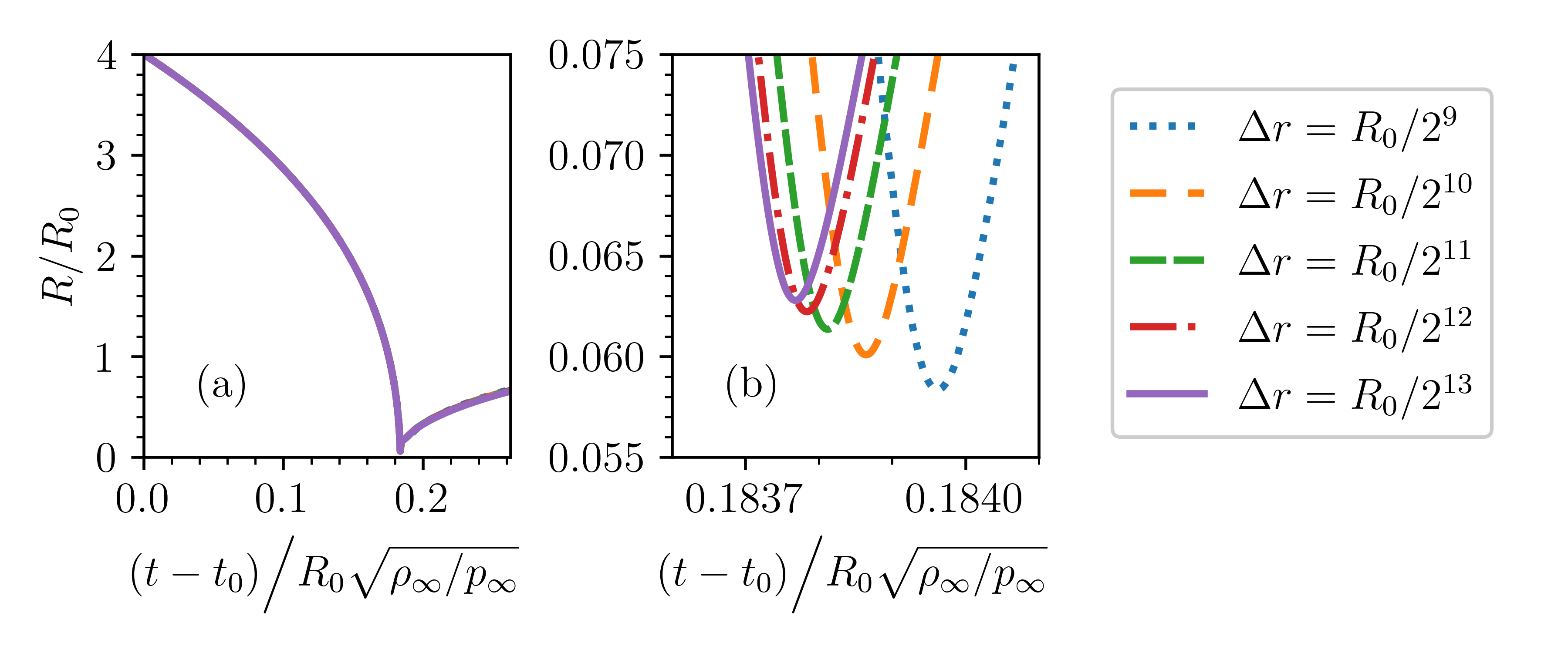} %for two-column figure, 6.69 inches x 8.25 inches. 
\caption{
Dependence of $R(t)$ on the resolution, parametrized by $\Delta r$ and $\Delta t_{\textrm{max}}$, during the strong collapse of an ideal xenon gas cavity surrounded by compressible water, with $R_m/R_0 = 20$.
The uniform bubble approximation for the gas motion is assumed, with the motion starting from $R(t_0) = R_m/5$.
The ambient parameters are $R_0 = 2.2\textrm{ }\upmu\textrm{m}$, $p_{\infty} = 1\textrm{ atm}$, and $T_{\infty} = 293\textrm{ K}$.
The numerical parameters are $\lambda = 8$ and $\Delta t_{\textrm{max}} = 5.5 \times 10^{-7} \times 2^9$ $\Delta r \sqrt{\rho_{\infty}/p_{\infty}}$.
In (b), the results shown in (a) are zoomed in near the minimum radius of the collapse.
}
\label{fig:water_convergence_in_dr}
\end{figure}

\begin{figure}
\includegraphics[width=4.8in, height=2.0in]{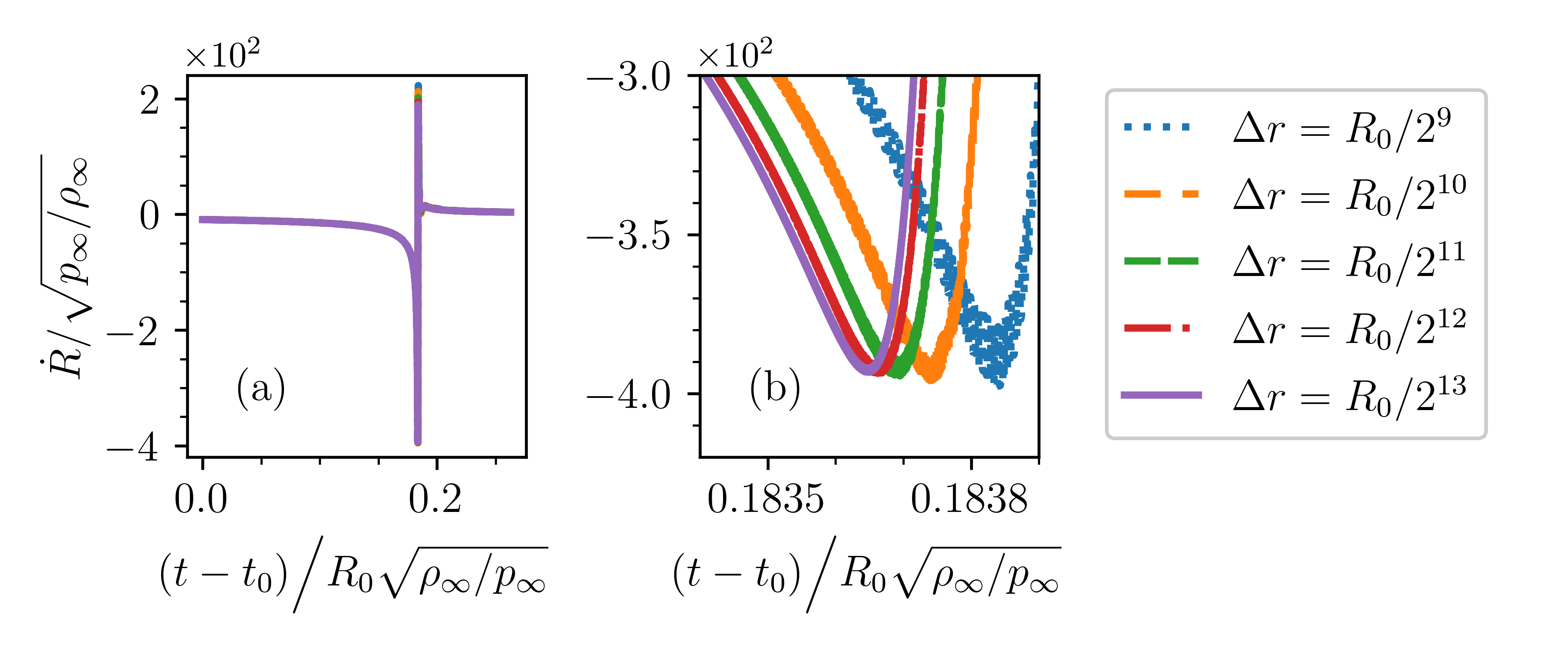} %for two-column figure, 6.69 inches x 8.25 inches. 
\caption{
Dependence of $\dot{R}(t)$ on the resolution, parametrized by $\Delta r$ and $\Delta t_{\textrm{max}}$, during the strong collapse of an ideal xenon gas cavity surrounded by compressible water, with $R_m/R_0 = 20$.
The uniform bubble approximation for the gas motion is assumed, with the motion starting from $R(t_0) = R_m/5$.
The ambient parameters are $R_0 = 2.2\textrm{ }\upmu\textrm{m}$, $p_{\infty} = 1\textrm{ atm}$, and $T_{\infty} = 293\textrm{ K}$.
The numerical parameters are $\lambda = 8$ and $\Delta t_{\textrm{max}} = 5.5 \times 10^{-7} \times 2^9$ $\Delta r \sqrt{\rho_{\infty}/p_{\infty}}$.
In (b), the results shown in (a) are zoomed in near the time of the maximum collapsing velocity.
}
\label{fig:water_convergence_in_dr_velocity}
\end{figure}

When analyzing the simultaneous convergence of $\Delta r$ and $\Delta t_{\textrm{max}}$, the initial values were taken as $\Delta r = R_0/2^9=R_0/512$ and $\Delta t_{\textrm{max}} = 5.5 \times 10^{-7}$ $R_0 \sqrt{\rho_{\infty}/p_{\infty}}$.
The results for $R(t)$ and $\dot{R}(t)$ are shown in Figs.~\ref{fig:water_convergence_in_dr} and \ref{fig:water_convergence_in_dr_velocity}, where it can be observed that the numerical solutions converge as spatial and temporal refinements increase.
For the least refined run, the computation took 19 minutes, while the most refined run, with $\Delta r = R_0/2^{13}=R_0/8192$ and $\Delta t_{\textrm{max}} = 5.5/16 \times 10^{-7}$ $R_0 \sqrt{\rho_{\infty}/p_{\infty}}$, took 3 days and 10 hours.
The observed scaling of computational time is approximately proportional to $(\Delta r)^{-2}$.
When comparing the two final levels of refinement with $\Delta r = R_0/2^{12}=R_0/4096$ and $\Delta r = R_0/2^{13}=R_0/8192$, the parameters $t_{\textrm{min}}$, $\dot{R}_{\textrm{max}}$, and $t_{\dot{R}_\textrm{max}}$ all differ by less than $0.1$\%, while $R_{\textrm{min}}$ changes by $0.9$\%.
Fig.~\ref{fig:water_convergence_in_dr_velocity} also shows that numerical oscillations in the velocity curves diminish with refinement. 
The results of the most refined run are also presented in Fig.~\ref{fig:comparison_of_water}.

\par

We also analyzed another strong collapse case with $R_m/R_0 = 20$, using the uniform bubble approximation and starting the calculation from $R_m/5$, where the ideal xenon gas bubble is surrounded by compressible liquid lithium.
The results and their convergence properties are analogous to those discussed earlier for water. 
The ambient parameters are taken as $R_0 = 2.2\textrm{ }\upmu\textrm{m}$, $p_{\infty} = 1\textrm{ atm}$, and $T_{\infty} = 454\textrm{ K}$.
Convergence with respect to $\lambda$ was performed with $\Delta r = R_0/2^9=R_0/512$ and $\Delta t_{\textrm{max}} = 2.7 \times 10^{-7}$ $R_0 \sqrt{\rho_{\infty}/p_{\infty}}$.
For $\lambda = 4$, the computation took approximately 14 minutes, while the largest value, $\lambda = 32$, ran for about 1 hour and 20 minutes.
The four parameters $R_{\textrm{min}}$, $t_{\textrm{min}}$, $\dot{R}_{\textrm{max}}$, and $t_{\dot{R}_\textrm{max}}$ varied by less than $10^{-6}\%$ between $\lambda = 16$ and $\lambda = 32$, so $\lambda = 16$ was chosen for the subsequent calculations.
\par

When analyzing the convergence of $\Delta r$ and $\Delta t_{\textrm{max}}$ simultaneously, the starting values were set to $\Delta r = R_0/2^9=R_0/512$ and $\Delta t_{\textrm{max}} = 2.7 \times 10^{-7}$ $R_0 \sqrt{\rho_{\infty}/p_{\infty}}$.
Again, we observe that the numerical solutions for the curves $R(t)$ and $\dot{R}(t)$ converge with increased spatial and temporal refinement.
For the least refined run, the computation took 1 hour and 37 minutes, while the most refined run, with $\Delta r = R_0/2^{12}=R_0/4096$ and $\Delta t_{\textrm{max}} = 2.7 / 8 \times 10^{-7}$ $R_0 \sqrt{\rho_{\infty}/p_{\infty}}$, took 6 days.
The observed scaling of the computational time is approximately proportional to $(\Delta r)^{-2}$.
When comparing the last two levels of refinement, with $\Delta r = R_0/2^{11}=R_0/2048$ and $\Delta r = R_0/2^{12}=R_0/4096$, the parameters $t_{\textrm{min}}$, $\dot{R}_{\textrm{max}}$, and $t_{\dot{R}_\textrm{max}}$ all differed by less than 0.1\%, while $R_{\textrm{min}}$ changed by 0.8\%.
The most refined run is also presented in Fig.~\ref{fig:comparison_of_lithium}.
\par

To understand the effects of compressibility, Figs.~\ref{fig:comparison_of_water} and \ref{fig:comparison_of_lithium} also show the solutions for the corresponding fluids within the incompressible approximation, where Eqs.~\eqref{radius_incompressible} and \eqref{pressure_gas_incompressible} are solved numerically starting from $R(t=0) = R_m$ and $\dot{R}(t=0) = 0$.
Additionally, the figures depict the asymptotic power-law regime, described by Eq.~\eqref{incompressible_Rayleigh_R_t}, that the full compressible solutions should approach as the radius decreases \cite{Hunter_1960, 10.1063/5.0160469}.
For a given $n$, the parameters $A$ and $t_c$ are computed by considering two points corresponding to $2R_{\textrm{min}}$ and $4R_{\textrm{min}}$ from the obtained compressible solution and assuming that these points lie on the power-law solution.
\par

\subsection{Shock wave emission from bubble collapse}

The desired upscaling of sonoluminescence may involve the use of opaque fluids such as liquid lithium. 
In these cases, the use of emitted light as a diagnostic for the strength of energy density concentration is limited.
However, a strong collapse will emit a shock wave into the surrounding fluid.
For instance, a cavitating bubble in water with $R_m/R_0 = 10$ has been shown to emit a shock wave that clearly registers a $3\textrm{ atm}$ signal on a needle hydrophone placed $1\textrm{ mm}$ from the cavity \cite{BARBER199765}.
\par

\begin{figure}
\includegraphics[width=5.2in, height=4.8in]{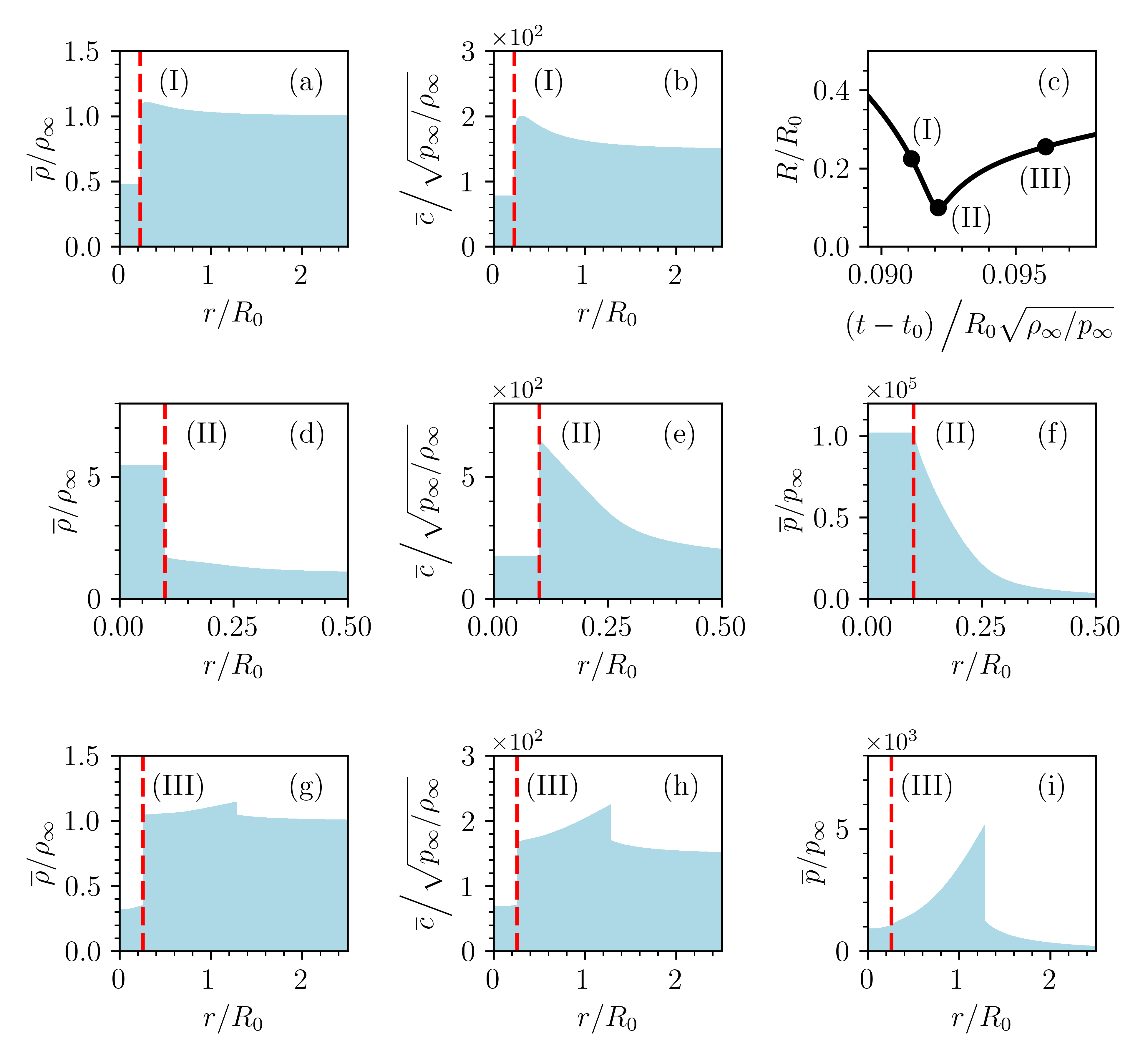} %for two-column figure, 6.69 inches x 8.25 inches. 
\caption{
Hydrodynamic behavior near the minimum radius for the strong collapse with $R_m/R_0 = 10$ of an ideal xenon gas cavity surrounded by compressible water.
The uniform bubble approximation for the gas motion is assumed, with the motion starting from $R(t_0) = R_m/5$.
Hydrodynamic profiles are shown before the uniform bubble approximation is emposed in a given time step.
The ambient parameters are $R_0 = 2.2\textrm{ }\upmu\textrm{m}$, $p_{\infty} = 1\textrm{ atm}$, and $T_{\infty} = 293\textrm{ K}$.
The numerical parameters are $\lambda = 8$, $\Delta r = R_0/2^{14}=R_0/16384$, and $\Delta t_{\textrm{max}} = 6.0/32 \times 10^{-7}$ $R_0 \sqrt{\rho_{\infty}/p_{\infty}}$.
The figure shows the solution at three times $t = 0.09112, 0.09212, 0.09612$, in units of $R_0 \sqrt{\rho_{\infty}/p_{\infty}}$, corresponding to (I), (II), (III), respectively. 
Figures (a), (d), (g) show profiles of the average mass density, $\overline{\rho}$; (b), (e), (h) show the average speed of sound, $\overline{c}$; and (f) and (i) show the average pressure, $\overline{p}$, with the red dashed line indicating the location of the cavity wall at each time. 
In (c), $R(t)$ is shown.
}
\label{fig:long_time_water_10_to_1}
\end{figure}

\begin{figure}
\includegraphics[width=5.2in, height=4.8in]{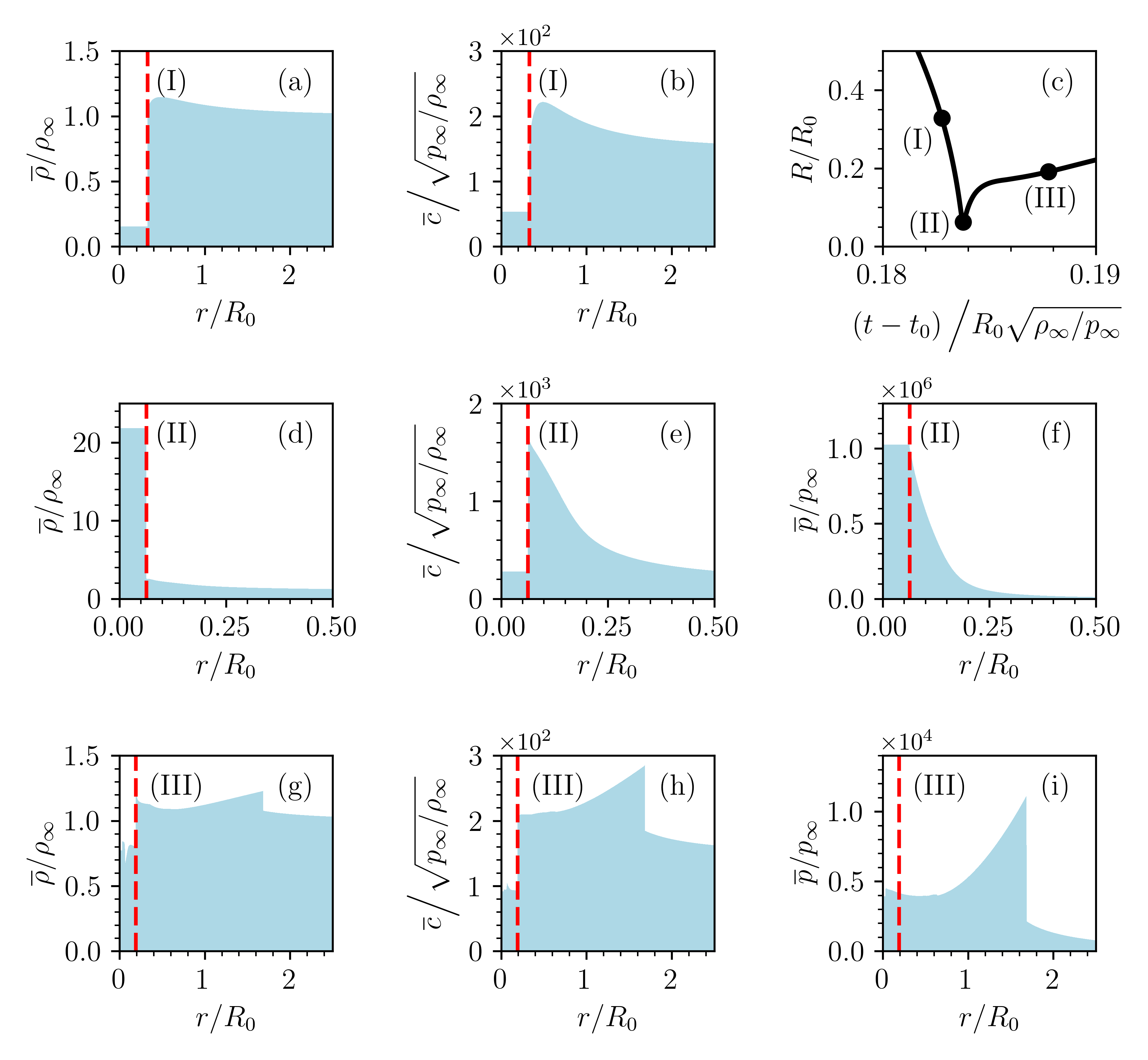} %for two-column figure, 6.69 inches x 8.25 inches. 
\caption{
Hydrodynamic behavior near the minimum radius for the strong collapse with $R_m/R_0 = 20$ of an ideal xenon gas cavity surrounded by compressible water.
The uniform bubble approximation for the gas motion is assumed, with the motion starting from $R(t_0) = R_m/5$.
Hydrodynamic profiles are shown before the uniform bubble approximation is emposed in a given time step.
The ambient parameters are $R_0 = 2.2\textrm{ }\upmu\textrm{m}$, $p_{\infty} = 1\textrm{ atm}$, and $T_{\infty} = 293\textrm{ K}$.
The numerical parameters are $\lambda = 8$, $\Delta r = R_0/2^{13}=R_0/8192$, and $\Delta t_{\textrm{max}} = 5.5/16 \times 10^{-7}$ $R_0 \sqrt{\rho_{\infty}/p_{\infty}}$.
The figure shows the solution at three times $t = 0.1828, 0.1838, 0.1878$, in units of $R_0 \sqrt{\rho_{\infty}/p_{\infty}}$, corresponding to (I), (II), (III), respectively. 
Figures (a), (d), (g) show profiles of the average mass density, $\overline{\rho}$; (b), (e), (h) show the average speed of sound, $\overline{c}$; and (f) and (i) show the average pressure, $\overline{p}$, with the red dashed line indicating the location of the cavity wall at each time. 
In (c), $R(t)$ is shown.
}
\label{fig:long_time_water_20_to_1}
\end{figure}

With the introduced approximations, it becomes possible to examine the hydrodynamic profiles of the compressible fluid near and after the point of minimum radius, where the bubble dynamics are the most extreme.
No additional constructions or ad hoc assumptions need to be added to the calculations to capture the emitted shock waves.
Examples of the mass density, speed of sound, and pressure profiles for the most refined simulation runs of water with $R_m/R_0 = 10$ and $R_m/R_0 = 20$ are shown in Figs.~\ref{fig:long_time_water_10_to_1} and \ref{fig:long_time_water_20_to_1}, respectively.
The figures show that we can resolve the formation of outgoing shock waves after the minimum radius.
These profiles are plotted at the time $4.0 \times 10^{-3}$~$R_0 \sqrt{\rho_{\infty}/p_{\infty}}$ after reaching the minimum radius.
At a given time from the singularity, the shock is about twice as strong for the stronger implosion.
At a distance of $3\textrm{ }\upmu\textrm{m}$, the shock wave emitted by a $R_m/R_0 = 10$ collapse has a strength of $4000~\textrm{atm}$.
A simple correction for the geometric spreading of the pressure amplitude suggests a strength of about $10~\textrm{atm}$ at $1~\textrm{mm}$.
As noted, this is within the range of experimental capabilities.
\par

\par

\par

\section{Numerical Results for Empty Cavity Collapse}

To further analyze the accuracy of the developed solver, we compare our solutions to the analytical asymptotic regime discussed in Refs. \onlinecite{Hunter_1960} and \onlinecite{10.1063/5.0160469}.
The asymptotic result holds in the limit as $R(t) \to 0$, so we now consider runs with an empty cavity, as this allows us to examine the behavior at much smaller radii.
To implement this, at each time step, we set uniform values of mass density and pressure within the bubble to be equal to $10^{-10}$ each, rather than using Eqs.~\eqref{density_incompressible} and \eqref{pressure_gas_incompressible}, which correspond to an ideal gas.
These chosen values are small enough for the gas in the bubble to behave as an empty cavity but large enough to avoid issues with division by zero.
The velocity profile in the cavity is still given by Eq.~\eqref{low_mach_velocity_in_the_gas}.
\par

Empty cavity calculations have been performed for both water and liquid lithium, and the obtained results are analogous.
Therefore, we now focus on the case of liquid lithium, where $R_m/R_0 = 20$ and the ambient parameters are $R_0 = 2.2\textrm{ }\upmu\textrm{m}$, $p_{\infty} = 1\textrm{ atm}$, and $T_{\infty} = 454\textrm{ K}$, as before.
The computation again starts from $R_m/5$.
In the case of empty cavity calculations, it is not possible to apply our previous algorithm for determining $\Delta t_{\textrm{max}}$ because there is no longer a minimum radius, and the velocity of the cavity wall, along with the pressure and speed of sound in the compressible fluid outside the cavity, increases indefinitely.
This leads to an arbitrarily small time step size from both adaptive time step schemes until it becomes unfeasible for the solver, resulting in the computation being aborted.
\par

To avoid this issue, we perform convergence analysis with respect to $\lambda$ using $\Delta r = R_0/2^9=R_0/512$ and $\Delta t_{\textrm{max}} = 2.7 \times 10^{-7}$ $R_0 \sqrt{\rho_{\infty}/p_{\infty}}$, as motivated by our liquid lithium calculations for a $R_m/R_0 = 20$ collapse of a bubble filled with ideal xenon gas.
For $\lambda = 4$, the computation took 13 minutes, while the largest value, $\lambda = 32$, ran for 1 hour and 10 minutes.
All four numerical parameters: $R_{\textrm{min}}$, $t_{\textrm{min}}$, $\dot{R}_{\textrm{max}}$, and $t_{\dot{R}_\textrm{max}}$, change by less than $2 \times 10^{-2} \%$ when going from $\lambda = 16$ to $\lambda = 32$.
Therefore, $\lambda = 16$ is selected for the subsequent calculations.
\par

When examining the convergence of simultaneous refinement of $\Delta r$ and $\Delta t_{\textrm{max}}$, we set the starting values to $\Delta r = R_0/2^9=R_0/512$ and $\Delta t_{\textrm{max}} = 2.7 \times 10^{-7}$ $R_0 \sqrt{\rho_{\infty}/p_{\infty}}$, again motivated by our previous simulations of strong collapse.
The least refined run took 1 hour and 26 minutes of computational time, while the most refined run, with $\Delta r = R_0/2^{12}=R_0/4096$ and $\Delta t_{\textrm{max}} = 2.7 / 8 \times 10^{-7}$ $R_0 \sqrt{\rho_{\infty}/p_{\infty}}$, took 4 days and 13 hours.
The observed scaling of the computational time remains approximately proportional to $(\Delta r)^{-2}$, similar to the case with the ideal gas inside the cavity.
Even between the last two levels of refinement, where $\Delta r = R_0/2^{11}=R_0/2048$ and $\Delta r = R_0/2^{12}=R_0/4096$, the calculations aborted at different final times, resulting in changes to $R_{\textrm{min}}$ and $\dot{R}_{\textrm{max}}$ of $12 \%$ and $11 \%$, respectively.
This highlights the difficulty of numerically capturing the behavior of the empty cavity at small radii.
Therefore, to analyze the convergence of the solutions, it is useful to compare them to the asymptotic theory.
\par

\begin{figure}
\includegraphics[width=3.84in, height=3.2in]{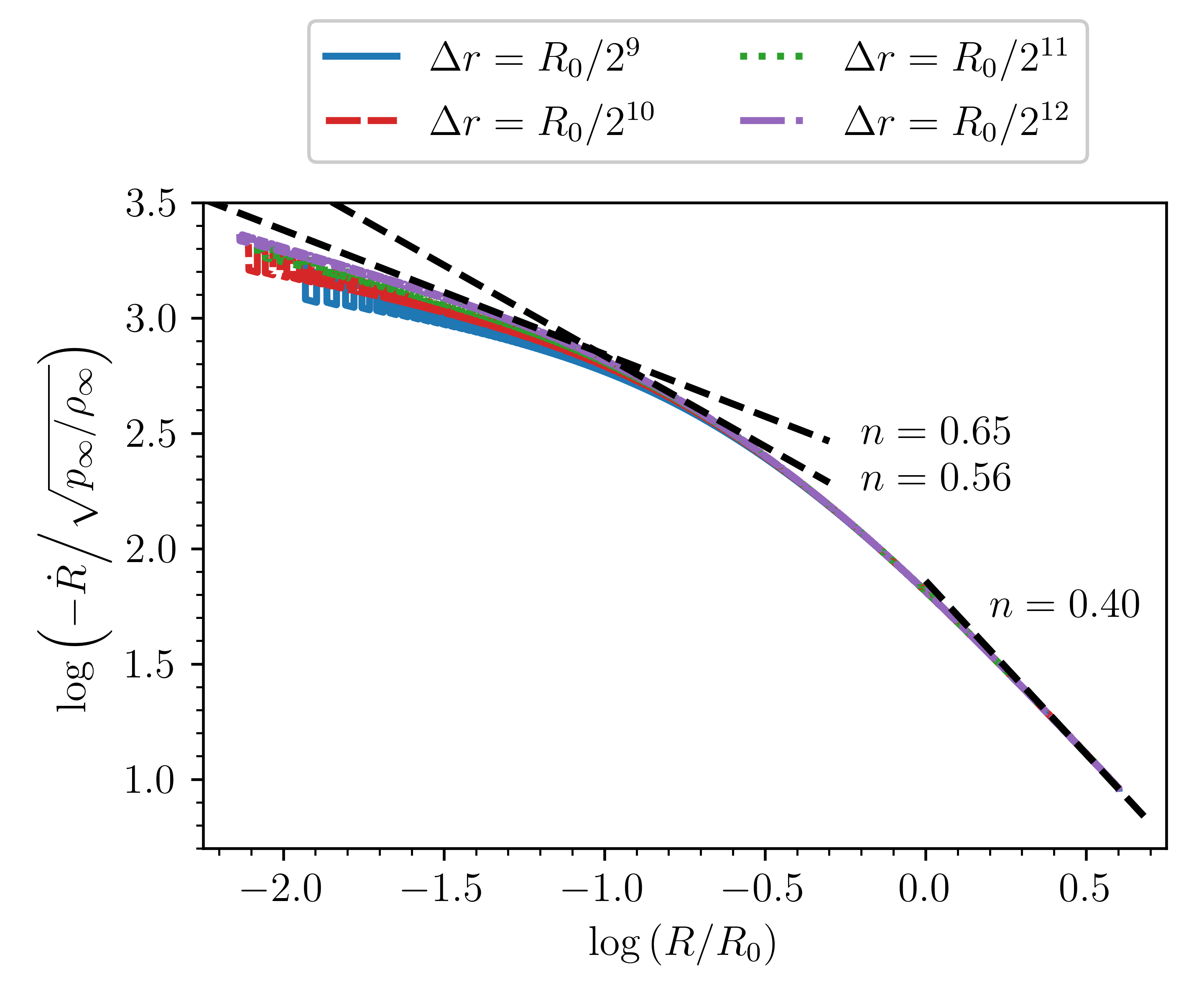} %for two-column figure, 6.69 inches x 8.25 inches. 
\caption{
The dependence of the empty cavity collapse on the resolution, parametrized by $\Delta r$ and $\Delta t_{\textrm{max}}$, when the cavity is surrounded by compressible liquid lithium with $R_m/R_0 = 20$.
The uniform bubble approximation for the cavity is assumed, and the motion starts from a radius $R_m/5$.
The ambient parameters are $R_0 = 2.2\textrm{ }\upmu\textrm{m}$, $p_{\infty} = 1\textrm{ atm}$, and $T_{\infty} = 454\textrm{ K}$.
The numerical parameters are $\lambda = 16$ and $\Delta t_{\textrm{max}} = 2.7 \times 10^{-7} \times 2^9 $ $\Delta r \sqrt{\rho_{\infty}/p_{\infty}}$.
Dashed lines show the power-law solution of the form given by Eq.~\eqref{incompressible_Rayleigh_R_t} for $n = 0.40, 0.56, 0.65$, corresponding to the asymptotic results of an incompressible fluid \cite{doi:10.1080/14786440808635681}, compressible water \cite{Hunter_1960}, and compressible liquid lithium \cite{10.1063/5.0160469}, respectively.
}
\label{fig:empty_cavity_lithium}
\end{figure}

To compare with the asymptotic theory that predicts the solution $R(t)$ to be of the form given in Eq.~\eqref{incompressible_Rayleigh_R_t} as $R(t) \to 0$, we note that this form satisfies $\dot{R} = -nA^{1/n}R^{(1-(1/n))}$, and so $\log(-\dot{R}) = \log (nA^{1/n}) + (1 - (1/n))\log R$. 
This implies that as $R(t) \to 0$, $\log(-\dot{R})$ should exhibit a linear dependence on $\log R$, with the slope determined by the appropriate value of $n$.
We plot the obtained solutions in these coordinates, as shown in Fig.~\ref{fig:empty_cavity_lithium}, where the dashed lines indicate slopes corresponding to various relevant values of $n$.
Indeed, we observe that the initial phase of the collapse closely follows the power-law solution with $n = 2/5$, which corresponds to the incompressible result \cite{doi:10.1080/14786440808635681}, as the velocity of the cavity is initially much smaller than the speed of sound in the surrounding fluid. 
As the radius of the cavity decreases, our results converge to the power-law solution with $n=0.65$, consistent with the asymptotic theory \cite{10.1063/5.0160469}.
Furthermore, we find that our result can be distinguished from the asymptotic result for other compressible fluids, such as water, which has $n = 0.56$ \cite{Hunter_1960}.
This outcome lends confidence to the proposed numerical method.

\section{Conclusion}

Our primary interest in cavity collapse is driven by the singular way it leads to a spontaneous concentration of energy density. 
We are particularly interested in the limitations on the extent to which energy can be focused, especially regarding the role of fluid compressibility in controlling this phenomenon.
To isolate the effects of compressibility, we ignore the influences of viscosity, surface tension, and mass transfer, which can be incorporated later if needed. 
A notable aspect of fluid compressibility is that, for adiabatic equations of state in the Tait-Murnaghan form, the radius of an empty cavity exhibits a finite-time singularity with an exponent that depends on the compressibility.
Capturing this exponent through computer simulations of the initial value problem for fluid motion is a crucial indicator of whether one has a physically relevant theory of strong collapse. 
This problem has been specifically formulated and developed here.
\par

Our approach builds upon an extension of Basilisk's open-source all-Mach solver, which utilizes a volume-of-fluid method to simulate compressible multiphase flows across all Mach number regimes.
Resolving the asymptotic exponent or simulating the parameter space of sonoluminescence requires a strong collapse, which we have found to be beyond the resolution of various studies on cavity collapse. 
Indeed, using initial conditions that are sufficiently violent to reach the asymptotic regime results in significant numerical anomalies in the simulation. 
For instance, consider a bubble starting from rest at its maximum radius.
After a very short time, a spurious vacuum region emerges near the bubble wall, even though the wall is moving inward.  
This, in turn, leads to extremely large, unphysical values for the computed speed of sound.
To overcome this issue, we have imposed the uniform bubble approximation, where at each step in the calculation, the gas inside the cavity has spatially uniform pressure and mass density profiles, consistent with the ideal gas equation of state. 
\par

Within our uniform bubble approximation, we have compared the collapse of a bubble surrounded by an incompressible fluid, a bubble in compressible water, and a bubble in compressible liquid lithium.
Our results indicate that the inclusion of compressibility effects slows down the collapse.
However, depending on the fluid, the incompressible regime can remain accurate for a considerable duration, with deviations occurring only near the minimum radius.
In each case, a strong collapse reaches an appropriate yet distinct asymptotic regime. 
The results for radius and velocity, as the most singular regions of motion are approached, have been simulated at increasing levels of resolution and found to converge to the analytically predicted asymptotes.
These simulations also facilitated the calculation of the strong outgoing pulses generated by the implosion.
These findings provide confidence that the effects of fluid compressibility have been adequately captured and that the method can be used to search for an optimal fluid in which the collapse of a gas-filled bubble would achieve the highest energy density. 
To pursue this goal, the interior gas will first be modeled using the van der Waals equation of state. 
The resulting interface dynamics will then be coupled with a gas modeled using molecular dynamics.
The objective is to determine the conditions under which the collapse leads to an imploding shock wave that reaches high temperatures. 
A key aim of this effort is to predict these temperatures, which we believe are significantly influenced by fluid compressibility.
\par

It remains to be seen whether there are alternative approaches to addressing the difficulties encountered when using the all-Mach solver for strong collapse.
An obvious approach is to conduct more refined calculations, but this leads to computational times that are impractical for exploring the sonoluminescence problem.
A potential improvement could involve transitioning from a fixed grid to an adaptive grid size, which might enable faster calculations while achieving a resolution sufficient to circumvent the aforementioned issues and accurately solve the full Euler equations for the gas within the cavity.

\begin{acknowledgments}
This research has been funded by AFOSR under FA9550-22-1-0425 (20223965).
The views, opinions, and/or findings expressed are those of the authors and should not be interpreted as representing the official views or policies of the Department of Defense or the U.S. Government.
S. J. R. also acknowledges the support of NSERC Canada (RGPIN 2022-03302).
We thank Y. Saade for many valuable discussions, and D. Fuster and M. Saini for the provided code.
\end{acknowledgments}

\bibliography{Basilisk_uniform_approximation}

\end{document}